\documentclass[twocolumn,pra,aps]{revtex4-1}

\usepackage{amssymb,amsmath}
\usepackage{graphicx}
\usepackage{comment}
\usepackage{color}
\usepackage[colorlinks,urlcolor=blue,citecolor=blue,linkcolor=blue]{hyperref}
\usepackage{cleveref}

\newcommand{\bra}[1]{\left<{#1}\right|}
\newcommand{\ket}[1]{\left|{#1}\right>}
\newcommand{\vect}[1]{{\mathbf #1}}

\newcommand{\down}{\downarrow}
\newcommand{\up}{\uparrow}
\renewcommand{\k}{{\bf k}}
\newcommand{\p}{{\bf p}}
\newcommand{\q}{{\bf q}}
\newcommand{\0}{{\bf 0}}

\newcommand{\ef}{\varepsilon_F}
\newcommand{\tf}{\tau_F}

\newcommand{\kf}{k_F}

\newcommand{\eq}{\epsilon_{\q}}

\newcommand{\ek}{\epsilon_{\k}}

\newcommand{\nn}{\nonumber}

\newcommand{\req}[1]{Eq.~\eqref{#1}}

\newcommand{\mim}{m_\text{im}}

\DeclareMathOperator{\re}{\mbox{Re}}
\DeclareMathOperator{\im}{\mbox{Im}}

\begin{document}

\title{Quantum dynamics of impurities coupled to a Fermi sea}

\author{Meera M. Parish}
\affiliation{School of Physics and Astronomy, Monash University, Victoria 3800, Australia.}

\author{Jesper Levinsen}
\affiliation{School of Physics and Astronomy, Monash University, Victoria 3800, Australia.}

\date{\today}

\begin{abstract}
We consider the dynamics of an impurity atom immersed in an ideal
Fermi gas at zero temperature.  We focus on the coherent quantum
evolution of the impurity following a quench to strong
impurity-fermion interactions, where the interactions are assumed to
be short range like in cold-atom experiments.  To approximately
model the many-body time evolution, we use a truncated basis method,
where at most two particle-hole excitations of the Fermi sea are
included.  When the system is initially non-interacting, we show
that our method exactly captures the short-time dynamics following
the quench, and we find that the overlap between initial and final
states displays a universal non-analytic dependence on time in this
limit.  We further demonstrate how our method can be used to compute
the impurity spectral function, as well as describe many-body
phenomena involving coupled impurity spin states, such as Rabi
oscillations in a medium or highly engineered quantum quenches.
\end{abstract}

\maketitle

\section{Introduction}

The coherent evolution of quantum many-body systems out of equilibrium
defines a new frontier in current research, and is of fundamental
importance to a number of fields, ranging from neutron stars to
electronic devices.  In fermionic systems, the investigation of
dynamics at the relevant time scale --- the Fermi time
$\tf=\hbar/\ef$, with $\ef$ the Fermi energy --- has recently become
available due to advances in the field of ultracold atoms. The
cold-atom system possesses a number of unique advantages over its
solid-state counterparts~\cite{Bloch2008}. Most notably, the
parameters of the governing models are precisely known and can be
rapidly changed; the cold-atom system is well isolated; and the real
time observation of coherent many-body dynamics is experimentally
accessible.  The Fermi time is typically in the microsecond range, in
stark contrast to the solid-state scenario where $\tf$ is shorter by
about 10 orders of magnitude due to the much lighter particles and
higher densities. The possibility of probing the coherent dynamics of
ultracold Fermi gases has stimulated a large theoretical effort to
understand interaction quenches in the crossover from
Bardeen-Cooper-Schrieffer (BCS) type superfluidity to a Bose-Einstein
condensate (BEC) of tightly bound pairs
\cite{Barankov2004,Andreev2004,Szymanska2005}.

A particularly clean realization of coherent dynamics on the Fermi
time scale is afforded by population imbalanced Fermi gases
\cite{Zwierlein2006,Zwierlein2006doo,Partridge2006pap,Schirotzek2009oof,Nascimbene2009,Liao:2010bh,Kohstall2012mac,Koschorreck2012aar,Bakr2016}. Here,
it is possible to investigate the dynamical response of a many-fermion
system to the sudden introduction of an impurity.  This response plays
a central role in important phenomena such as the orthogonality
catastrophe \cite{Nozieres1969sit}.  A recent experiment employed
Ramsey interferometry on heavy $^{40}$K impurity atoms immersed in a
$^6$Li Fermi sea, revealing both the real-time formation of impurity
quasiparticles as well as the interference between attractive and
repulsive polaron branches \cite{Cetina2016}.  The Ramsey protocol
provides a direct measure of the time-dependent overlap function at
time $t$~\cite{Goold2011oca,knap2012tdi}
\begin{align}
S(t) &=\left<\psi_0\right|e^{i\hat H_0t}e^{-i\hat H_\text{int}t}\left|\psi_0\right>,
\label{eq:1}
\end{align}
where $\ket{\psi_0}$ is the initial non-interacting state of the total
system, $t=0$ defines the point where the impurity starts interacting
with the Fermi sea, and $\hat H_0$ and $\hat H_\text{int}$ correspond
to the Hamiltonians in the absence and presence of interactions,
respectively. As such, Ramsey interferometry provides detailed
information on the time evolution of the impurity wave function.
Throughout this article, we focus on the purely quantum evolution at
\textit{zero} temperature, and we work in units where $\hbar$, the
Boltzmann constant $k_B$ and the volume are all set to 1.

The dynamical response of a strongly interacting quantum many-body
system is a challenge to determine theoretically since the
interactions cannot be treated perturbatively.  Here, we present a
theoretical approach to determine the coherent impurity dynamics based
on truncating the Hilbert space of impurity wave functions at a fixed
number of particle-hole excitations of the Fermi sea. As we
demonstrate, this truncated basis method (TBM) allows us to capture
the Ramsey response exactly at times $t\ll \tf$, where two-body
correlations dominate. We also consider the challenging scenario of an
infinitely heavy (static) impurity, where one has the orthogonality
catastrophe \cite{Nozieres1969sit}, and the overlap in
Eq.~\eqref{eq:1} exhibits a power-law decay at long times arising from
the multiple low-energy excitations of the Fermi sea.  In this case,
we show that the TBM provides results that are essentially exact to
several $\tf$. Hence, the TBM likely provides a near exact solution up
to several $\tf$ for coherent impurity dynamics in the strongly
interacting regime, even when the impurity mass is finite.
Furthermore, we argue that the TBM also captures the long-time Ramsey
response in cases where the attractive polaron
\cite{Chevy2006upd,Prokofev2008} is well-defined.

A key result of the present work is the exact short-time evolution of
the Ramsey response, which is dominated by two-body physics for
$t\ll \tf$. In the case where the short-range interaction of the
$\ket{\up}$ impurity with the Fermi sea is described by a single
parameter, the scattering length $a$, the Ramsey response takes the
form
\begin{align} 
S(t) &\simeq 1 - \frac{8e^{-i\pi/4}(m/m_r)^{3/2}}{9\pi^{3/2}}  \left(\frac{t}{\tf}\right)^{\frac32}.
\label{eq:shortintro}
\end{align}
Here, $m$ ($m_\text{im}$) is the mass of a majority (impurity)
particle, and $m_r=m m_\text{im}/(m+m_\text{im})$ is the reduced
mass. Note that Eq.~\eqref{eq:shortintro} does not depend on the
scattering length and, furthermore, it does not display the short-time
behavior expected from a simple expansion of the time evolution
operators, where we have $1-S(t)\propto t^2$. Such a quadratic
dependence on time is also expected for a Loschmidt echo
\cite{Loschmidt1876udz}, which is defined as a time-dependent wave
function overlap similar to Eq.~\eqref{eq:1} and which yields
information about an imperfect time-reversal procedure applied to a
quantum system. Instead, the non-analytic behavior of $S(t)$ is a
direct consequence of the renormalization of the contact
interactions. For resonances where the effective range $r_\text{eff}$
greatly exceeds the van der Waals range of the interatomic
interactions, $r_\text{eff}$ must be taken into account in the short
time evolution. In this case, we find
\begin{align}
S(t) &\! \simeq\!1 \!- \!\frac{(m/m_r)^2}{3\pi k_FR^*}\left(\frac{t}{\tf}\right)^{2}
\!+\!\frac{16e^{i\pi/4}
       (m/m_r)^{5/2}}{45\pi^{3/2} (k_FR^*)^2}  \left(\frac{t}{\tf}\right)^{\frac{5}{2}}\!,
\label{eq:shortrs2intro}
\end{align}
where $k_F=\sqrt{2m\ef}$ is the Fermi momentum and we define the
positive range parameter $R^*=-r_\text{eff}/2$. Again, the Ramsey
response is independent of scattering length, and while the leading
order contribution has the expected form of a Loschmidt echo, the next
order correction is non-analytic.

The TBM provides us with a framework in which impurity dynamics can be
explored systematically. To illustrate this point, we apply it to two
scenarios of coherent impurity dynamics beyond the Ramsey response:
Rabi oscillations between quasiparticle branches, and the dynamical
preparation of strongly interacting quantum states. We also show how
the TBM allows the straightforward calculation of the impurity
spectral function.

The paper is organized as follows. In Sec.~\ref{sec:model}, we
describe the model Hamiltonian, while in Sec.~\ref{sec:TBM} we outline
the truncated basis method. In Sec.~\ref{sec:response} we present our
results for the Ramsey response, including the analytic short and
long-time behavior, as well as for the impurity spectral function.
Sections \ref{sec:rabi} and \ref{sec:stateprep} discuss, respectively,
Rabi oscillations and how the initial quantum state can be modified.
Section \ref{sec:2ph} then examines the role played by multiple
particle-hole excitations, focussing for simplicity on a static
impurity. We conclude in Sec.~\ref{sec:conc}.

\section{Model}
\label{sec:model}
In the following, we consider the dynamics of a single impurity
immersed in a Fermi gas. For this purpose, it is convenient to
consider two impurity spin states, $\sigma=\down,\up$, of which one
($\up$) is strongly interacting with the Fermi sea, while the other
($\down$) is non-interacting.  To model interactions, we employ a
two-channel Hamiltonian. Restricting ourselves at first to the part of
the Hamiltonian describing the interacting $\up$ impurity state and
the medium, we have
\begin{align}
\hat{H}_{\rm int} =& \sum_\k \epsilon_{\k,\rm{im}} \hat c^\dag_{\k\up}\hat c_{\k\up}
+\sum_{\vect{k}} \epsilon_{\vect{k}}  \hat f^\dag_{\vect{k}}\hat
           f_{\vect{k}}
+ \sum_\vect{k} \left[\epsilon_{\vect{k},M}+\nu 
\right]
\hat d_\vect{k}^\dag \hat d_\vect{k} \nonumber \\ &\hspace{-4mm}+ g\sum_{\vect{k},\vect{q}} %\chi(\vect{k})
\left(\hat d^\dag_\vect{q} \hat f_{\q/2+\k}\hat c_{\q/2-\k,\up}
+\hat c^\dag_{\q/2-\vect{k},\up} \hat f^\dag_{\q/2+\vect{k}} \hat d_\vect{q} \right).
\label{eq:twochannel}
\end{align}
The first line of Eq.~\eqref{eq:twochannel} corresponds to the
non-interacting Hamiltonian $\hat{H}_0$, where
$\hat c^\dag_{\k\sigma}$ ($\hat c_{\k\sigma}$) creates (annihilates)
an impurity particle with momentum $\k$, spin $\sigma$, mass $\mim$,
and single particle energy
$\epsilon_{\k,\text{im}}=\frac{k^2}{2\mim}$. Likewise, the operators
$\hat f^\dag_\k$ and $\hat f_\k$ respectively create and annihilate a
majority fermion with momentum $\k$, mass $m$, and single particle
energy $\epsilon_\k=\frac{k^2}{2m}$.  The spin-$\up$ impurity
interacts with the fermions by forming a closed channel molecule
described by the creation and annihilation operators $\hat d^\dag_\k$
and $\hat d_\k$ with momentum $\k$, single-particle energy
$\epsilon_{\k,\text{M}}=\frac{k^2}{2M}$, and mass $M=m+\mim$. The
detuning of this closed channel molecule from the impurity-fermion
scattering threshold is denoted $\nu$. The interaction --- second line
of \req{eq:twochannel} --- has a coupling strength $g$ for relative
momenta with magnitude $|\k| < \Lambda$, where $\Lambda$ is a UV
cut-off.

Using standard techniques (see, e.g., Ref.~\cite{Gurarie2007}), we
relate the bare interaction parameters $g$, $\Lambda$, and $\nu$ to
renormalized quantities by calculating the low-energy spin-$\up$
impurity-fermion scattering amplitude at a relative momentum $\k$
within the model \eqref{eq:twochannel}. We then compare the resulting
expression with the standard low-energy expansion of the scattering
amplitude
\begin{align}
f(\k)=-\frac1{a^{-1}-\frac12r_\text{eff}k^2+ik},
\end{align}
where $a$ and $r_\text{eff}$ are the scattering length and effective
range, respectively. This procedure yields the scattering length $a$ through
\begin{align}
\frac{m_r}{2\pi a} &= -\frac{\nu}{g^2} +
\sum_{\vect{k}}^\Lambda\frac{1}{\epsilon_{\vect{k}}+\epsilon_{\vect{k},\rm{im}}},
\label{eq:a}
\end{align}
In particular, we see how the model allows us to tune the scattering
length to resonance, $1/a=0$. For resonances where $|r_\text{eff}|$
greatly exceeds the range of the bare interaction, $r_\text{eff}$ is
negative and we instead define the range parameter
\cite{Petrov2004tbp}
\begin{align}
R^* = -r_\text{eff}/2=\frac{\pi}{m_r^2 g^2}.
\label{eq:rs}
\end{align}
We emphasize that the model \eqref{eq:twochannel} reduces to the
commonly used single-channel model with $R^*=0$ by taking
$g,\nu\to \infty$ in such a way that
$\nu/g^2=m_r\Lambda/\pi^2-m_r/(2\pi a)$.

The presence of the auxiliary $\down$ state enables one to probe
impurity dynamics starting from a non-interacting state. For instance,
interactions between the impurity and the medium can suddenly be
switched on by using a radiofrequency (rf) pulse which couples the
$\down$ and $\up$ impurity states. The total Hamiltonian is then
$\hat H=\hat H_{\rm int}+\hat H_{\rm aux}$ with
\begin{align}
\hat H_{\rm aux} = & \sum_\k (\epsilon_{\k,{\rm im}}+\delta)\hat c^\dag_{\k\down}\hat c_{\k\down}
\nn \\ & + \frac{\Omega_0}{2i}\sum_\k (e^{i\varphi}\hat c^\dag_{\k\down} \hat c_{\k\up} - e^{-i\varphi}\hat c^\dag_{\k\up} \hat c_{\k\down}).
\label{eq:Haux}
\end{align}
Here, $\Omega_0$ and $\varphi$ are the Rabi frequency and phase of the
rf pulse, respectively, and $\delta\equiv \omega-\omega_0$ is the
detuning of the rf pulse with frequency $\omega$ from the bare
$\down$--$\up$ transition frequency, $\omega_0$.  Note that we have
applied the rotating wave approximation, assuming
$|\delta/(\omega+\omega_0)|\ll1$. For the remainder of this
manuscript, we set $\omega_0$ to zero.

\section{Truncated basis method} \label{sec:TBM} To formulate our
approach to the many-body dynamics, we start from the time-dependent
variational principle, which is applicable to any many-body system and
is not limited to impurity dynamics.  Here, we wish to determine the
time evolution of an approximate variational wave function $\psi(t)$
that best describes that of the actual system. To this end, we
consider the action of the operator
$\hat{\epsilon} = i\partial_t - \mathcal{H}$, where $\mathcal{H}$ is
the Hamiltonian that governs the dynamics of the system.  Clearly, if
$\psi$ were the exact wave function, then $\hat{\epsilon} \psi =0$.
More generally, if we know $\psi(t)$ at time $t$ and we wish to
approximately determine $\psi(t+\delta t)$, we must minimise the
``error'' quantity~\cite{McLachlan64}
\begin{align} \label{eq:error}
 \Delta =\int |\hat{\epsilon}\psi(t)|^2 dV,
\end{align}
with respect to the unknown function $\partial_t\psi$, where the above
integral is over all space. There are also other formulations of the
time-dependent variational principle that give equivalent
results~\cite{PhysRevE.51.5688}.

In what follows, we will consider wave functions of the form:
$\ket{\psi} = \sum_j \gamma_j \ket{j}$, where $\{\ket{j} \}$
represents a subset of a complete orthonormal set of states. Within
this truncated basis, Eq.~\eqref{eq:error} becomes
\begin{align} \notag
\Delta = & \ i  \sum_j \left(\dot{\gamma}^*_j \bra{j} \mathcal{H} \ket{\psi} - \dot{\gamma}_j \bra{\psi} \mathcal{H} \ket{j} \right)  
\\ 
&+ \sum_{j,l}  \bra{j}l \rangle \dot{\gamma}^*_j \dot{\gamma}_l +\bra{\psi}\mathcal{H}^2 \ket{\psi},
\end{align}
where $\dot{\gamma}_j = \frac{d\gamma_j}{dt}$.  Imposing the condition
$\frac{\partial \Delta}{\partial\dot{\gamma}^*_j} = 0$ then gives
\begin{align}
\bra{j} \hat{\epsilon} \ket{\psi} = 0.
\end{align}
Exploiting the orthonormality of the basis states,
$\bra{j}l \rangle = \delta_{jl}$, finally yields the equations of
motion:
\begin{align}\label{eq:motion}
i  \frac{d\gamma_j}{dt} = \sum_l  \bra{j} \mathcal{H} \ket{l} \gamma_l \equiv \sum_l\mathcal{H}_{jl} \gamma_l.
\end{align}
Note that the norm of the wave function is preserved in this case
since we have $\bra{\psi} \hat{\epsilon} \ket{\psi} = 0$, i.e.:
\begin{align}
  \frac{d}{dt} \bra{\psi} \psi \rangle 
  & = i\left(\bra{\psi} \mathcal{H} \ket{\psi} -  \bra{\psi} \mathcal{H} \ket{\psi}\right) = 0.
\end{align}
Equivalently, we can see this from the fact that the time evolution
operator within this subspace is unitary.

\subsection{General solution for a quench} \label{sec:vari} To
determine the approximate time evolution of a system, one must in
general solve the set of coupled differential equations
\eqref{eq:motion} directly. However, the situation simplifies when the
system evolves under a time-independent Hamiltonian. This includes the
scenario where there is an abrupt change in the parameters of the
Hamiltonian at some time $t_0$, i.e., the system undergoes a
\textit{quench}, which is the focus of this paper.

In this case, one proceeds by solving for the eigenstates of the
projected Hamiltonian $\mathcal{H}_{jl}$, using the equations for the
energy $E$:
\begin{align}
    E \gamma_j = \sum_l\mathcal{H}_{jl} \gamma_l,
\label{eq:variation}
\end{align}
and then expanding the system's wave function $\ket{\psi(t)}$ in terms
of these eigenstates.  To be concrete, suppose we start from some
initial state $\ket{\psi(0)}$, and we instantaneously ``turn on'' the
effect of the Hamiltonian $\mathcal{H}$ at time $t=0$.  The resulting
wave function is
\begin{align}
\ket{\psi(t)}&= e^{-i\mathcal{H}t}\left|\psi(0)\right>  \simeq \sum_n  \bra{\phi_n} \psi(0) \rangle e^{-iE_n t} \ket{\phi_n},
\end{align}
where $\ket{\phi_n}$ are the eigenstates within the $\{\ket{j}\}$
subspace, and $E_n$ are the corresponding eigenenergies. Note that
this implicitly assumes that the eigenstates are orthogonal, but this
is guaranteed from the fact that the Hamiltonian is Hermitian in this
subspace.

\subsection{Impurity wave function}
For the specific case of an impurity interacting with a Fermi medium,
we consider wave functions of the form
\begin{subequations}
\begin{align}
    \ket{\psi}& = \ket{\psi_\up}+\ket{\psi_\down},
\end{align}
where
\begin{align}
\ket{\psi_\up} 
= & \left[ \alpha_{0} \hat c^\dag_{\mathbf{0}\up} 
 + \sum_\vect{q}  \alpha_\vect{q} \hat d^\dag_\vect{q} \hat f_{\vect{q}}   +  \sum_{\vect{k}, \vect{q}}
\alpha_{\vect{k}\vect{q}}\hat c^\dag_{\vect{q}- \vect{k}\up}
\hat f^\dag_{\vect{k}} \hat f_{\vect{q}}\right] \left| \rm{FS}\right>,  \\
\ket{\psi_\down} 
= & \left[ \beta_{0} \hat c^\dag_{\0\down} 
 +  \sum_{\vect{k}, \vect{q}}
\beta_{\vect{k}\vect{q}}\hat c^\dag_{\vect{q}- \vect{k}\down}
\hat f^\dag_{\vect{k}} \hat f_{\vect{q}}\right] \left| \rm{FS}\right>.
\end{align}
\label{eq:tbm}
\end{subequations}
Here,
$\left| \rm{FS}\right> = \prod_{|\vect{q}|<k_F}\hat f^\dag_{\vect{q}}
\ket{0}$
describes the non-interacting Fermi sea with energy
$E_0 = \sum_{|\vect{q}|<k_F}\epsilon_{\vect{q}}$. Thus, we require
that $|\vect{k}|>k_F$ and $|\vect{q}|<k_F$ for particle and hole
excitations, respectively, in the impurity wave functions.  We have
implicitly assumed the impurity to have zero momentum, which is a good
approximation in the limit of a small impurity density and zero
temperature.  The wave functions illustrate how the impurity can
excite particles out of the Fermi sea leaving holes behind, and in
both spin channels we truncate the wave function at one particle-hole
excitation.  The difference between $\ket{\psi_\up}$ and
$\ket{\psi_\down}$ arises from the fact that the interacting spin
$\up$ impurity can bind a majority fermion to form a closed-channel
molecule.  Note that similar wave functions have been used to describe
the equilibrium properties of an impurity in a Fermi gas: For the
ground state, this includes the attractive quasiparticle
\cite{Chevy2006upd,CombescotLoboChevy2007,combescot2008nso} and the
transition to a dressed dimer state
\cite{combescot2009,punk2009,Mora2009gso}, as well as the transition
to a trimer ground state \cite{Mathy2011tma}. Such wave functions have
also been used to describe the metastable upper branch, i.e., the
repulsive polaron \cite{Cui2010}.

When diagonalizing the Hamiltonian within the truncated basis spanned
by states of the form \eqref{eq:tbm}, we note that the interaction
part of the Hamiltonian, Eq.~\eqref{eq:twochannel}, still contains
both the bare coupling $g$ and the detuning $\nu$. While $g$ is
related to $R^*$ via Eq.~\eqref{eq:rs}, it is not possible to
immediately trade the detuning for the renormalized interaction
parameter, the scattering length, as the momentum cut-off cannot be
taken to infinity. Thus, in practice we apply the following procedure:
For a given $a$ and $R^*$, we first choose a momentum cut-off
$\Lambda$ and adjust $\nu$ according to Eq.~\eqref{eq:a} to obtain the
desired scattering length. Next, we increase the momentum grid to
obtain convergent results at the chosen momentum cut-off, repeating
this step for increasing cut-off to ensure convergence of the final
result.

\section{Coherent impurity dynamics following a quench \label{sec:response}}
We now consider Ramsey interferometry and its relation to the impurity
spectral function.  To provide illustrations of the method in this
section, we focus on equal masses $m_{\rm im} = m$.  However, the
approach can straightforwardly be extended to a mass imbalanced
mixture \cite{Cetina2016}.

\subsection{Dynamical response to an interaction quench}

\begin{figure}[th]
        \centering
                \includegraphics[width=0.95\columnwidth]{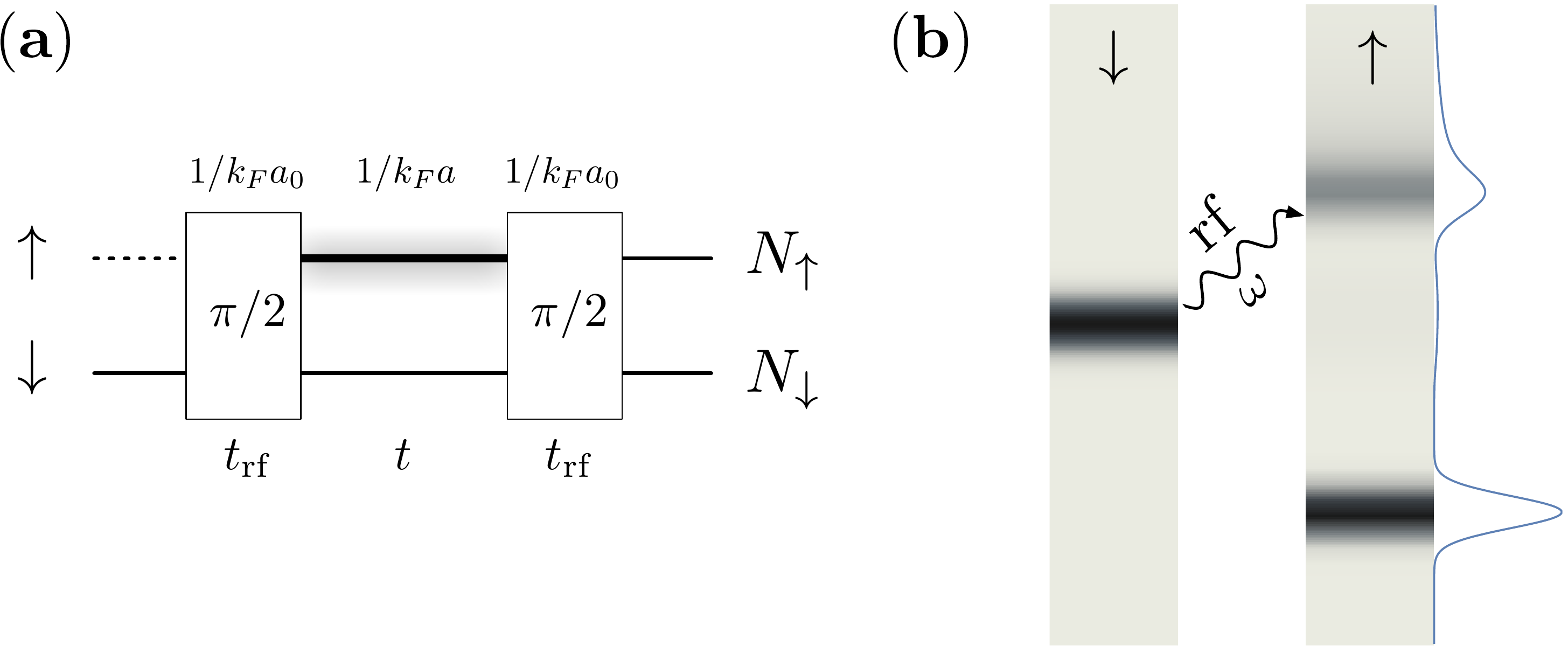}
                \caption{Illustration of experimental procedures used
                  to probe impurities in a Fermi gas. (a) In Ramsey
                  interferometry, the impurities are initially in the
                  non-interacting $\ket{\down}$ state.  At time $t=0$,
                  they are in a superposition of $\ket{\down}$ and
                  $\ket{\up}$ states following a $\pi/2$ rf pulse of
                  duration $t_{\rm rf}$.  After a variable time $t$
                  and a second $\pi/2$ rf pulse, the number of
                  particles, $N_\sigma$, in the two impurity spin
                  states is measured as a function of the phase of the
                  second pulse. The interaction between the $\up$
                  impurity state and the majority fermions is
                  characterized by the interaction parameter
                  $1/k_Fa_0$ during the rf pulses and $1/k_Fa$ during
                  the evolution time. (b) In inverse rf spectroscopy,
                  the spectral response of spin $\down$ impurities to
                  an rf pulse is measured as a function of
                  frequency~$\omega$ relative to the bare
                  $\down$-$\up$ transition frequency (vertical axis).}
        \label{fig:expt} 
\end{figure}

We first consider the scenario where an impurity, initially in the
non-interacting spin-$\down$ state, is suddenly coupled to an
interacting spin-$\up$ state by an rf pulse.  The many-body response
to a rapidly introduced impurity into the Fermi gas can be probed by
means of Ramsey interferometry~\cite{Goold2011oca,knap2012tdi}, as
illustrated in Fig.~\ref{fig:expt}(a): Following an initial $\pi/2$ rf
pulse, which creates a superposition of the impurity in $\down$ and
$\up$ spin states, the system evolves under the interacting
Hamiltonian for a time $t$, after which a second $\pi/2$ rf pulse is
applied. For simplicity, in this section we consider a `perfect
quench' where no interactions take place during the rf pulses, and
thus at time $t=0$ the impurities are in an equal superposition
$\frac1{\sqrt{2}}(\ket{\down}+\ket{\up})$.  In this case, a
measurement of the impurity population difference at the end of the
Ramsey procedure yields~\footnote{Here we have assumed that all the
  closed channel molecules have been converted into spin-up atoms at
  the time of the measurement, which is reasonable if there is a
  sufficient delay after the second rf pulse.}
\begin{align} \label{eq:Ndiff}
    \frac{N_\up - N_\down}{N_\up + N_\down} = - {\rm Re} \left[ e^{i \varphi_{\rm rf}} S(t) \right] + n_d,
\end{align}
where $\varphi_\text{rf}$ is the phase of the second rf pulse with
respect to the first, $n_d$ is the fraction of closed channel
molecules at time $t$, and we have the overlap between interacting and
non-interacting states
\begin{align}\label{eq:St}
S(t) &=\left<\psi_0(t)|\psi_\text{int}(t)\right>=e^{iE_0t}\left<\psi_0\right|e^{-i\hat H_{\rm int}t}\left|\psi_0\right>,
\end{align}
where $\ket{\psi_0} \equiv \hat c^\dag_{\mathbf{0}\up} \ket{{\rm FS}}$
and $\hat H_0\ket{\psi_0}=E_0\ket{\psi_0}$. By varying the relative
phase $\varphi_\text{rf}$, one can thus access both the amplitude and
phase of $S(t)$.

\begin{figure*}[th]
        \centering
        \includegraphics[width=2.1\columnwidth]{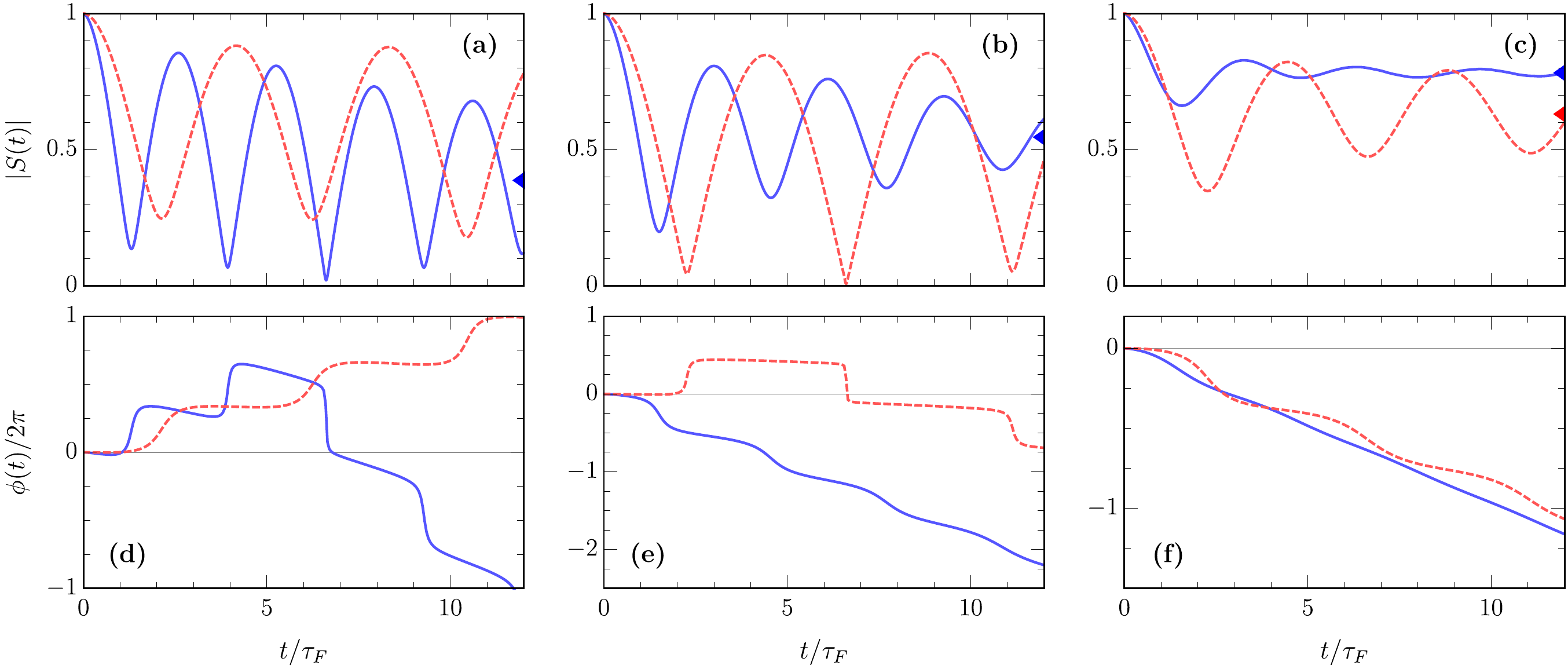}
        \caption{The amplitude (top) and phase (bottom) of the Ramsey
          signal. We have taken the interaction parameter $1/k_Fa$ to
          be (a,d) $0.73$, (b,e) $0.48$, and (c,f) 0. The range
          parameter is $k_FR^*=0$ (solid) and $k_FR^*=1$ (dashed). The
          diamonds indicate the long-time limit of the Ramsey response
          in the cases where the attractive polaron is the ground
          state. The phase for $k_F R^* = 0$ in (f) is well
          approximated by $\phi(t) \simeq E_{\rm att} t$, where the
          attractive polaron energy $E_{\rm att} \simeq$ $-0.607 \ef$
          \cite{Chevy2006upd}.  }
        \label{fig:Soft} 
\end{figure*}

\begin{comment}
erga0048 = -1.13917; Za0048 = 0.545939;
erga00 = -0.606648; Za00 = 0.782169;
ergr0073 = 0.689994; Zr0073 = 0.57575;
ergr0048 = 0.871104; Zr0048 = 0.405067;

erga1073 = -1.00919; Za1073 = 0.334248;
erga1048 = -0.814415; Za1048 = 0.433998;
erga10 = -0.540476; Za10 = 0.630822;
ergr1073 = 0.489157; Zr1073 = 0.615948;
ergr1048 = 0.591751; Zr1048 = 0.515259;
ergr10 = 0.852091; Zr10 = 0.3;
\end{comment}

According to the variational approach outlined in Sec.~\ref{sec:vari},
we can determine an approximate Ramsey response $S(t)$ by
diagonalizing the Hamiltonian within the subspace of wave functions of
the form \eqref{eq:tbm}. In the perfect quench scenario, we only need
to consider the decoupled spin-up part of the Hamiltonian,
$\hat H_{\rm int}$; thus we obtain the set of equations
\cite{Trefzger2012}:
\begin{align}
  (E-E_0)\alpha_0 & =g\sum_\q\alpha_\q, \nn \\
  (E-E_0)\alpha_\q & = (\epsilon_{\q,\text{M}}-\epsilon_\q+\nu)\alpha_\q+g\alpha_0+g\sum_\k \alpha_{\k\q}, \nn \\
  (E-E_0)\alpha_{\k\q} & =(\epsilon_{\q-\k,\text{im}}+\ek-\eq)\alpha_{\k\q}
                         +g\alpha_\q.
\label{eq:tbmeqs}
\end{align}
Solving these coupled equations yields the set of eigenstates
$\ket{\phi_j}$ with corresponding energies $E_j$. We then obtain for
the Ramsey response
\begin{align}
S(t)\simeq \sum_{j} \left|\left<\psi_0 | \phi_j \right>\right|^2 e^{-i(E_j-E_0)t}.
\label{eq:STBM}
\end{align}
This expression has a natural interpretation. Up to a trivial phase,
the contribution from the state $\ket{\phi_j}$ rotates at an angular
frequency $E_j$, while the magnitude of the contribution is the
squared overlap with the non-interacting ground state, i.e., the
residue of $j$'th state:
$Z_j\equiv \left|\left<\psi_0 | \phi_j \right>\right|^2$.

The time evolution of the impurity after an interaction quench is
clearly intrinsically connected to the structure of its energy
spectrum. As we discuss in more detail in Sec.~\ref{sec:spectral}, the
spectrum can contain well-defined quasiparticle states (the attractive
and repulsive polarons) as well as a broad continuum of many-body
states which have a vanishing overlap with the non-interacting
system. The interference of these different states is, in general,
expected to generate damped coherent oscillations in $|S(t)|$ as a
function of time.

Figure~\ref{fig:Soft} shows both the amplitude and the phase of
$S(t)\equiv |S(t)|e^{-i\phi(t)}$ for different values of the
interaction and the range parameter.  The slope of the phase $\phi(t)$
gives an indication of whether the energies in the impurity spectrum
are predominantly positive or negative.  In general, we observe that
the amplitude near $t=0$ is characterized by an initial descent that
is independent of scattering length and is only sensitive to $R^*$.
The quantum evolution then displays oscillations on a time scale which
is set by the Fermi time $\tf=1/\ef$.  In (c,f), the dynamics is
dominated by the attractive ground-state polaron, while for stronger
attraction, the evolution can feature roughly equal contributions from
the attractive and repulsive branches of the system, thus leading to
pronounced oscillations in $|S(t)|$ and $\phi(t)$.

\begin{figure}[th]
        \centering
        \includegraphics[width=.9\columnwidth]{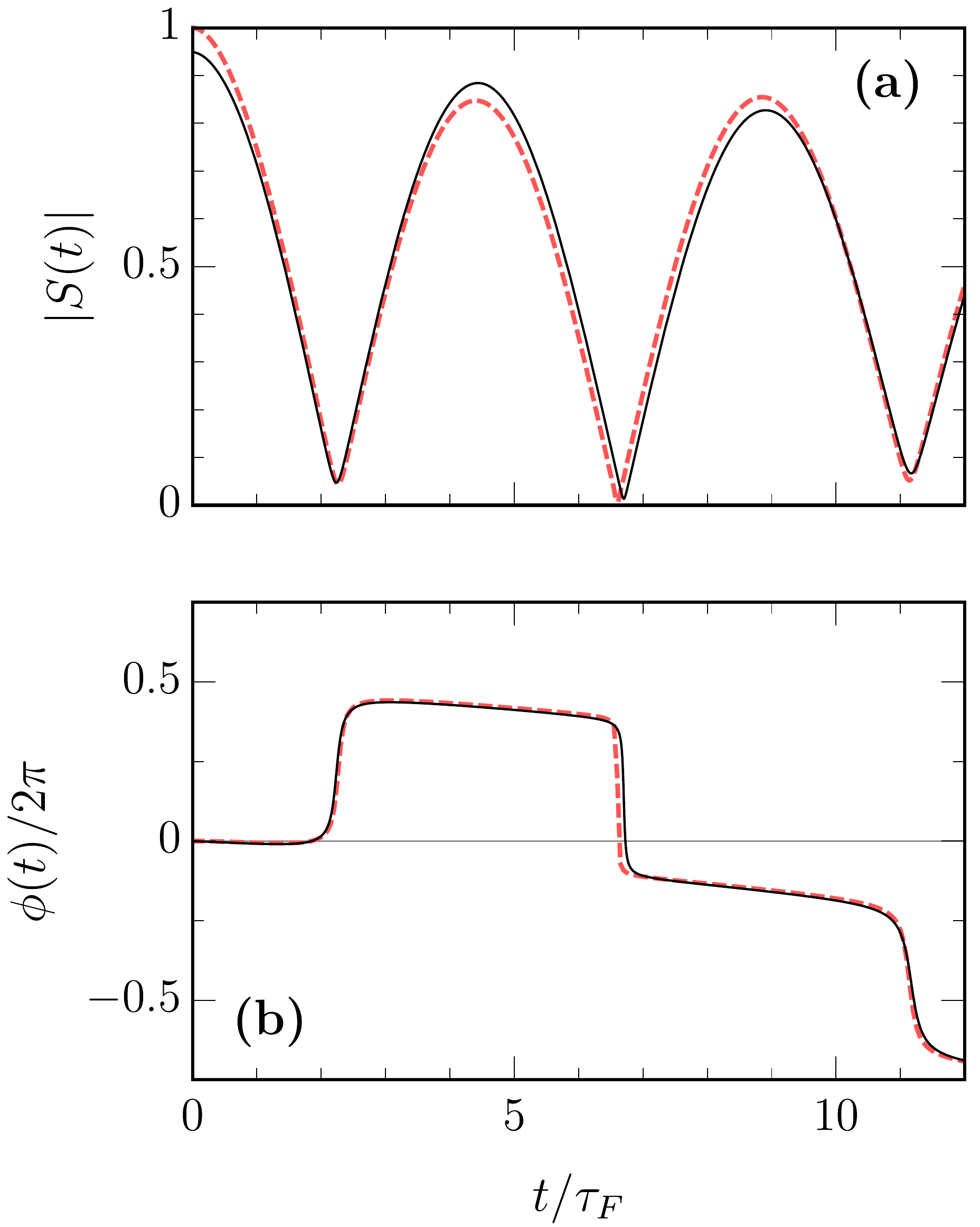}
        \caption{(a) Amplitude and (b) phase of the Ramsey signal
          (dashed lines) together with the corresponding results from
          the approximation Eq.~\eqref{eq:SoftApprox} (thin solid
          lines) for $1/k_Fa=0.48$ and $k_FR^*=1$. For these
          interaction parameters, we have $Z_\text{att}\approx0.55$
          and $Z_\text{rep}\approx0.41$ \cite{Massignan2012}. We have
          added a small imaginary part $i\ef/33$ to the repulsive
          polaron energy to model the finite quasiparticle lifetime.}
        \label{fig:SoftApprox} 
\end{figure}

To quantity this further, we assume the attractive and repulsive
branches are well-defined polaron quasiparticles, and consider the
regime where both of their residues --- $Z_\text{att}$ and
$Z_\text{rep}$, respectively --- are close to 1/2. We can then analyse
the Ramsey response in terms of the interference between the two
polarons. Assuming that we can ignore the contribution from the
continuum of states, we approximate the Ramsey response by
\begin{align}
S(t) \simeq Z_\text{att} \ e^{-i E_\text{att} t} + Z_\text{rep} \ e^{-i E_\text{rep} t} 
\ ,
\label{eq:SoftApprox}
\end{align}
with $E_\text{att}$ ($E_\text{rep}$) the attractive (repulsive)
polaron energy. As illustrated in Fig.~\ref{fig:SoftApprox}, this
approximation describes the response -- in particular, the period of
the beats -- very well.  Thus, the effect of $1/k_F a$ and $k_F R^*$
on the dynamics may be simply estimated from their effect on the
quasiparticle energies and residues.  Sharp jumps in the phase
accompany the regions where the amplitude approaches zero, and the
direction of these jumps is the only feature of the dynamics that
sensitively depends on the quasiparticle lifetime. Otherwise, we may
assume both quasiparticles to be infinitely long lived.  The validity
of the approximation \eqref{eq:SoftApprox} hinges on the small residue
of the states in the continuum that lies between the attractive and
repulsive peaks. This feature is also observed in recent diagrammatic
Monte Carlo calculations \cite{Goulko2016}.  However, note that this
is not necessarily true for arbitrary impurity mass, and indeed we
find a larger weight in the continuum for a heavy
impurity~\cite{Cetina2016}.

The behavior of the Ramsey response at times greatly exceeding $\tf$
is determined by the ground state of the impurity problem. If the
attractive polaron is the ground state, this implies that there is a
well-defined quasiparticle peak of zero width in the impurity spectral
function. Hence, while all contributions to $S(t)$ in
Eq.~\eqref{eq:STBM} originating from the higher-lying continuum of
states interfere destructively and thus dephase, this single term
becomes dominant. Therefore, in this limit, $|S(t)|%\stackrel{t\gg\tf}
{\to} Z_\text{att}$
and $\phi(t)\to E_\text{att}t$, corresponding to the formation of the
attractive polaron.  Since wave functions of the form (\ref{eq:tbm})
provide a good approximation to the residue and energy of the
attractive polaron \cite{combescot2009}, we therefore expect that the
TBM will accurately describe the long-time behavior of the Ramsey
response for sufficiently weak interaction strengths where the
attractive polaron is the ground state.

\subsection{Spectral function} \label{sec:spectral} We now discuss how
the dynamical response of the impurity to an interaction quench is
related to the \emph{spectral} response obtained using inverse rf
spectroscopy.  In the latter case, we start with impurities in the
non-interacting spin $\down$ state, and then apply an rf pulse that
couples the two impurity spin states, as described by the Rabi term in
the Hamiltonian --- see Eq.~\eqref{eq:Haux}. Assuming a weak pulse,
$\Omega_0 \ll \ef$, such that it can be treated within linear response
theory, the fraction of atoms transferred is directly proportional to
the impurity spectral function. The protocol is illustrated in
Fig.~\ref{fig:expt}(b). The impurity spectral function has been
measured in several ultracold atom experiments
\cite{Schirotzek2009oof,Kohstall2012mac,Koschorreck2012aar}. Theoretically,
it has previously been treated within the renormalization
group~\cite{Schmidt2011}, in diagrammatic Monte
Carlo~\cite{Goulko2016}, and within a T matrix
approach~\cite{Massignan2011}.  The latter approach includes two-body
correlations in the impurity wave function systematically, and is
equivalent to the TBM calculation of the spectral function with one
particle-hole excitation.  However, the TBM is easier to extend to
other types of impurity dynamics and to higher order correlations, as
we show in Secs.~\ref{sec:rabi}-\ref{sec:2ph}.

\begin{figure}[th]
        \centering
        \includegraphics[width=.8\columnwidth]{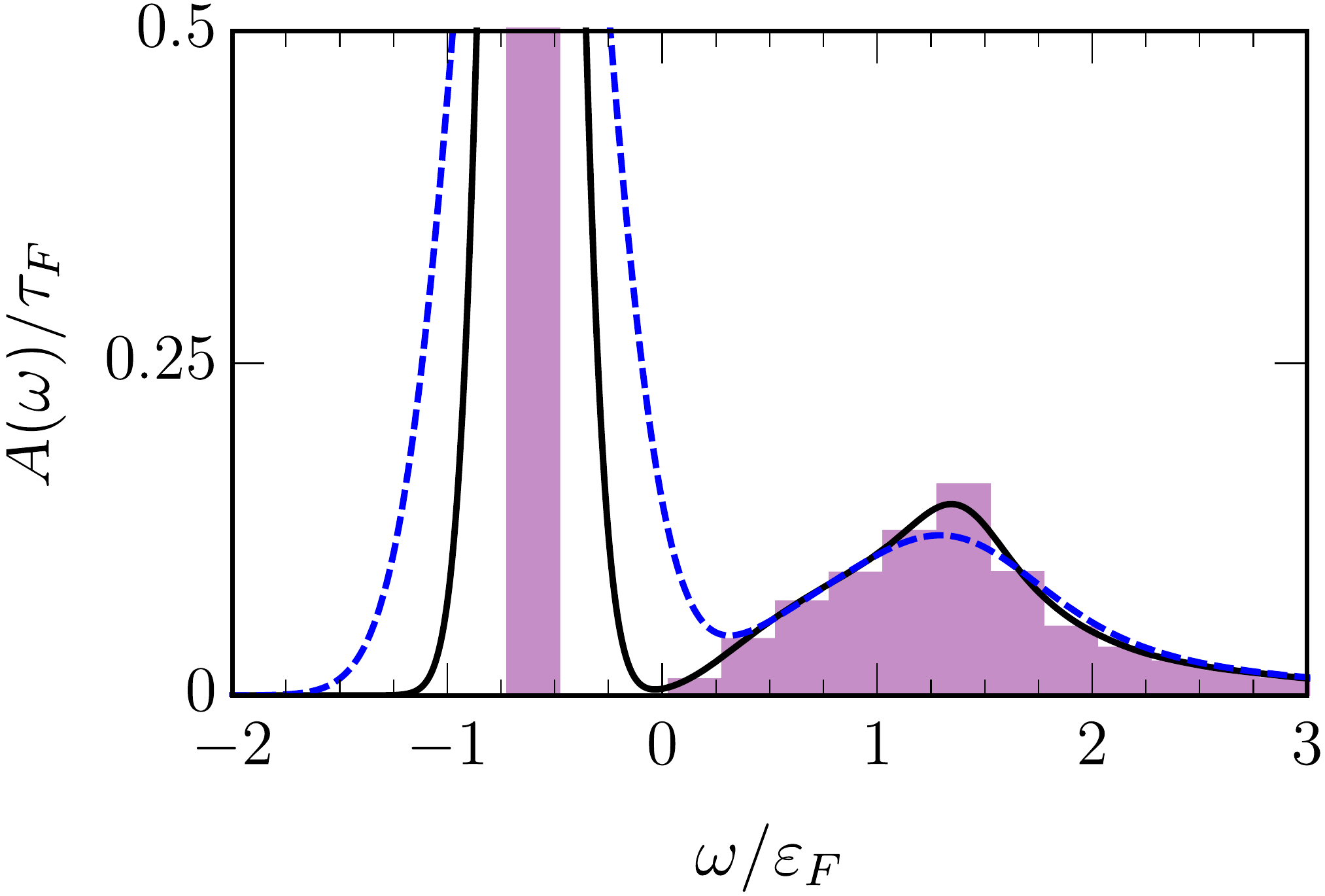}
        \caption{Illustration of the decomposition of the spectral
          function at $1/k_Fa=k_FR^*=0$. The solid line is the
          convolved spectral function $I(\omega)$ according to
          Eq.~\eqref{eq:I} with a Gaussian width of $\sigma=0.15E_F$
          (solid) and $\sigma=0.3E_F$ (dashed). The bars show
          $A(\omega)$ calculated according to Eq.~\eqref{eq:A}, where
          the eigenvalues have been binned and the height of each bin
          set to
          $\sum_{j\in \text{bin}}|\left<\psi_0|\phi_j\right>|^2$.  }
        \label{fig:decomp} 
\end{figure}

\begin{figure*}[th]
        \centering
        \includegraphics[width=\textwidth]{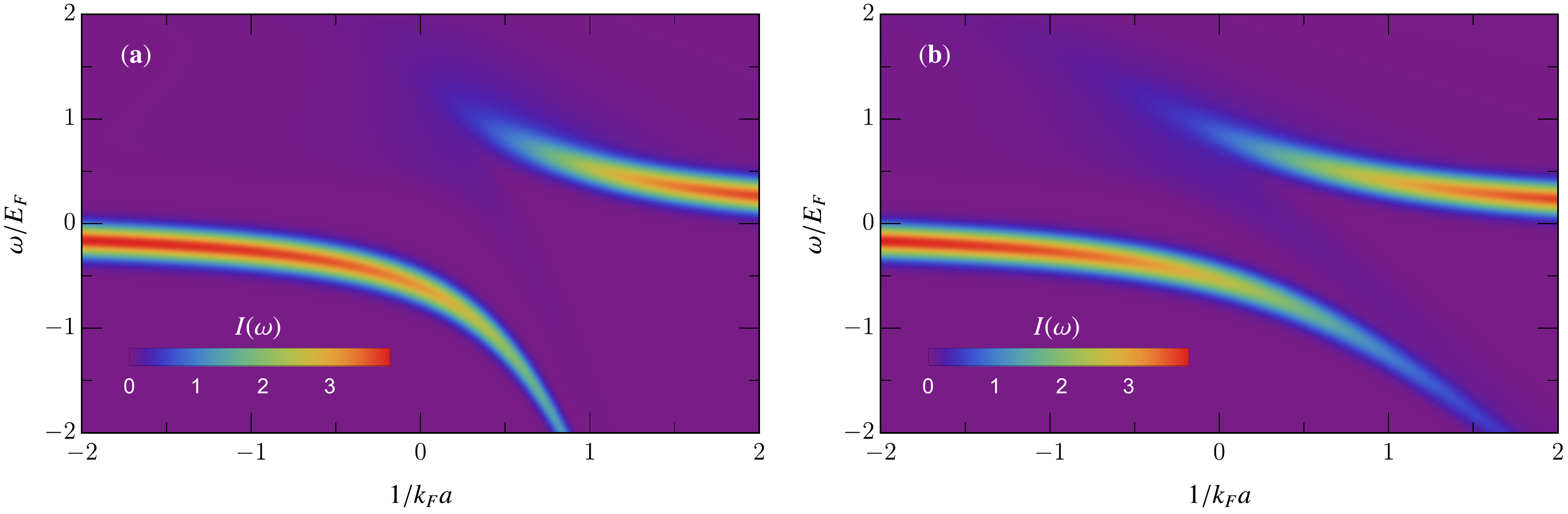}
        \caption{Spectral function $I(\omega)$ calculated within the
          TBM with one particle-hole excitation for (a) $k_FR^*=0$ and
          (b) $k_FR^*=1$. The Gaussian width is taken to be
          $\sigma=0.1E_F$.  The dressed dimer state is the ground
          state rather than the attractive polaron when
          $1/\kf a \gtrsim 0.9$ in (a) and $1/\kf a \gtrsim 0.4$ in
          (b).}
        \label{fig:spectra} 
\end{figure*}

For a perfect quench in the dynamical problem, the Ramsey response
\eqref{eq:St} corresponds to the overlap between the time-evolved
interacting and non-interacting states of the system. The spectral
function $A(\omega)$ is then obtained from the Fourier transform of
$S(t)$ \cite{mahan1990mpp}:
\begin{align} \label{eq:AwSt}
A(\omega) & =  \re \int^\infty_0 \frac{dt}{\pi} e^{i\omega t} S(t).
\end{align}
This clearly illustrates the close connection between the energy
spectrum and the dynamical response of the system to a quench of the
system parameters.

Using Eq.~\eqref{eq:STBM}, we can find an approximate spectral
function within the TBM,
\begin{align}
\label{eq:A}
  A(\omega) & \simeq \sum_{j} \left|\left<\psi_0 | \phi_j \right>\right|^2 \underbrace{\int^\infty_{-\infty} \frac{dt}{2\pi} e^{i\omega t} e^{-i(E_j-E_0)t}} _{\delta(\omega-E_j+E_0)} \ .
\end{align}
The Dirac delta function is easier to handle in the discretized basis
if we first convolve $A(\omega)$ with a Gaussian of width $\sigma$:
\begin{align} \notag
I(\omega) & = \int_{-\infty}^\infty d\omega' A(\omega-\omega') g(\omega') \\  \label{eq:I}
& = \sum_{j} \left|\left<\psi_0 | \phi_j \right>\right|^2 g(\omega - E_j+E_0),
\end{align}
where 
\begin{align}
g(\omega) = \frac{1}{\sqrt{2\pi}\sigma} e^{-\omega^2/2\sigma^2}.
\end{align}
Indeed, such a convolution mirrors experiment, where the spectral
response is determined using rf pulses of a finite duration and hence
a non-zero width in frequency space. This width can typically be well
approximated by a Gaussian.

In Fig.~\ref{fig:decomp}, we illustrate the idea behind the method:
First we evaluate the raw spectrum of energy eigenvalues and
corresponding residues, which yields a large number of discrete peaks
of variable heights. The convolved spectral function, $I(\omega)$, on
the other hand, is a smooth function of frequency and is what would be
observed in experiment. Such a spectral convolution is easier to
generate using the TBM compared to the standard T matrix
approach~\cite{Massignan2011}.

We show the results of this procedure in Fig.~\ref{fig:spectra} for
two values of $k_F R^*$. We see that the spectrum in both cases is
dominated by the attractive and repulsive polaron quasiparticles at
positive and negative energy, respectively. In between, there is a
broad continuum of states which all have a very small wave function
overlap with the non-interacting impurity state. In particular, once
$1/k_Fa\gtrsim1$, the spectral weight of the continuum is essentially
negligible, as was also observed in Ref.~\cite{Goulko2016}.

The main effect of the range parameter $R^*$ is to shift the energies
of the polaron branches closer to zero, especially in the unitary
regime, and to increase the lifetime of the repulsive polaron such
that it can be well-defined even on the attractive side of the
resonance \cite{Massignan2013}.  It also affects the character of the
impurity ground state: With increasing $1/k_Fa$, the impurity
eventually undergoes a sharp transition from an attractive polaron to
a dressed dimer~\cite{Mora2009gso,punk2009,combescot2009}, and this
transition occurs at lower $1/k_Fa$ for larger $k_F R^*$.  However,
this is not captured by the TBM with one particle-hole excitation,
since the attractive polaron always remains the ground state at this
level of truncation.

\subsection{Short-time dynamics}
\label{sec:ultrafast}
We now turn to the limiting behaviour of $S(t)$ at short times when
$t\lesssim \tau_F$. Away from resonance, the results presented in the
following furthermore require $t\ll2m_ra^2$ when $|k_Fa|\ll1$. We
start by formally Taylor expanding the time evolution operator in
Eq.~\eqref{eq:St}, which yields:
\begin{align}
S(t) \simeq 1 -i \left<\psi_0\right|\delta\hat H\left|\psi_0\right> t - \frac{1}{2}\left<\psi_0\right|(\delta\hat H)^2\left|\psi_0\right>t^2,
\end{align}
where $\delta\hat H = \hat H_{\rm int} - E_0$.  Using
Eq.~\eqref{eq:AwSt}, we see that the first term yields the usual sum
rule for the spectral function:
\begin{align}
\int_{-\infty}^\infty d\omega\,A(\omega)  & = 1.
\end{align}
For the second term, using the two-channel Hamiltonian
\eqref{eq:twochannel}, we simply obtain:
\begin{align}
\left<\psi_0\right|\delta\hat H\left|\psi_0\right> = g 
\bra{\rm{FS}} \hat{c}_{\vect{0}\up} \sum_{\vect{q}} \hat d^\dag_\vect{q} \hat f_{\vect{q}} \ket{\rm{FS}} = 0.
\end{align}
Thus, the leading order behavior of $S(t)$ is determined by the last
term
\begin{align}
\left<\psi_0\right|(\delta\hat H)^2\left|\psi_0\right> = g^2 %\frac{g^2}{V} 
\sum_{|\vect{q}|<k_F} = \frac{g^2k_F^3}{6\pi^2},
\end{align}
which finally gives:
\begin{align}
S(t) \simeq 1 - \frac{k_F^3 t^2}{12\pi R^* m_r^2}.
\label{eq:shortrs}
\end{align} 
These results yield an additional set of sum rules for the spectral
function of the impurity (see also Ref.~\cite{mueller2013}):
\begin{align}
  \int_{-\infty}^\infty d\omega \, \omega
  A(\omega)  & = 0, 
  \\ \label{eq:2ndmoment}
\int_{-\infty}^\infty d\omega \, \omega^2 A(\omega)  & = \frac{k_F^3}{6\pi R^* m_r^2}.
\end{align}
Note that in the limit of a broad resonance where $g\to \infty$,
Eq.~\eqref{eq:2ndmoment} diverges and there is no well-defined
short-time parabolic decay of $|S(t)|$. As such, it does not resemble
the initial decay of the Loschmidt echo expected for quantum systems
in this case.  Indeed, even when $g$ is finite, we find that terms
involving higher powers of $\delta\hat H$ are divergent, e.g., for the
next order term, we obtain
\begin{align}
\left<\psi_0\right|(\delta\hat H)^3\left|\psi_0\right> = g^2 
\sum_{|\vect{q}|<k_F} \left( \epsilon_{\vect{q},M} - \epsilon_{\vect{q},Li} + \nu \right),
\end{align}
which clearly diverges for short-range interactions since
$\nu \sim \Lambda \to \infty$.

The origin of these divergences is the non-analytic behavior of the
many-body wave function when the distance between the impurity and a
majority fermion goes to zero.  One thus needs to isolate the
high-frequency behavior of $A(\omega)$ in order to address the
short-time dynamics in the presence of short-range interactions.

From the Green's function $G(\omega)$ for an impurity at zero
momentum, we have $A(\omega) = -\im[G(\omega)]/\pi$, where we can, in
turn, write the Green's function in terms of the self energy
$\Sigma(\omega)$:
\begin{align}
G(\omega) = \left[\omega - \Sigma(\omega) \right]^{-1}.
\end{align}
In the limit $\omega \to \infty$, we can neglect non-trivial effects
of the Fermi medium, i.e., the self-energy is dominated by two-body
scattering, giving:
\begin{align}
\Sigma(\omega) \simeq n \mathcal{T}(\omega) ,
\end{align}
where the medium density $n = \frac{k_F^3}{6\pi^2}$ and the two-body T
matrix for $\omega >0$ is
\begin{align}
\mathcal{T}(\omega) = \frac{2\pi}{m_r} \left( a^{-1} + 2m_r \omega R^* + i\sqrt{2m_r \omega} \right)^{-1}.
\end{align}
Thus, the high-frequency limit of the spectral function is contained
in the expression
\begin{align} \label{eq:Alim}
A(\omega) \simeq 
\frac{k_F^3}{3\pi^2 \sqrt{2 m_r^3} \, \omega^{5/2}}
\frac{1}{1+2R^*/a+2 m_r \omega {R^*}^2}. 
\end{align}

Focussing first on the case $R^* = 0$, the leading order correction to
$S(t)$ in the limit $t\to 0$ can be determined from the integral:
\begin{align}
\int^\infty_{\tilde{\omega}} d\omega A(\omega) \left(e^{-i\omega t} - 1 +i\omega t \right),
\end{align}
where $\tilde{\omega}$ is a large frequency scale that can be sent to
infinity at the end of the calculation.  Inserting
Eq.~\eqref{eq:Alim}, we finally obtain
\begin{align} \notag
S(t) & \simeq 1 + \frac{k_F^3 t^{3/2}}{3\pi^2 \sqrt{2 m_r^3}}   \lim_{t \to 0}\left[ \int^\infty_{\tilde{\omega} t} d\omega \frac{\left(e^{-i\omega} -1 + i\omega  \right)}{\omega^{5/2}} \right] \\
&= 1 - \frac{2 \sqrt{\pi} (1-i) k_F^3}{9\pi^2 \sqrt{m_r^3}} \ t^{3/2}
\label{eq:short}
\end{align}
for $t>0$, with $S(-t)=S^*(t)$. We have thus succeeded in deriving
Eq.~\eqref{eq:shortintro} from the introduction. This universal,
non-analytic expression for the short-time behavior is a key result of
this paper.

For the case where $R^* > 0$, the non-analytic behavior appears in the
next order term of $S(t)$, which can be obtained from the integral:
\begin{align} \notag &\lim_{t\to 0} \int_{\tilde{\omega}}^\infty
  d\omega A(\omega) \left(e^{-i\omega t} - 1 +i\omega t + \frac{1}{2}
    \omega^2 t^2 \right) %\right]
  \\ \notag &= \frac{k_F^3 t^{5/2}}{6\pi^2 {R^*}^2\sqrt{2 m_r^5}}
              \lim_{t\to 0} \left[ \int^\infty_{\tilde{\omega} t}
              d\omega \frac{\left(e^{-i\omega} -1 + i\omega +
              \frac{1}{2} \omega^2 \right)}{\omega^{7/2}} \right].
\end{align}
Evaluating the integral, we thus obtain for the short-time expansion
of $S(t)$:
\begin{align}
S(t) & \simeq 1 - \frac{k_F^3 t^2}{12\pi R^* m_r^2}+\frac{2\sqrt{\pi} (1+i)
       k_F^3}{45\pi^2 {R^*}^2\sqrt{m_r^5}}  \ t^{5/2},
       \label{eq:shortrs2}
\end{align}
which demonstrates Eq.~\eqref{eq:shortrs2intro} from the
introduction. At first glance, one might expect the short-time
behavior of the Ramsey response (or high-frequency tail of the
spectral function) to be connected to the Tan contact
\cite{Tan2008eoa}.  However, we emphasize that the Tan contact governs
the large-frequency behavior of the occupied spectral function in the
equilibrium system, not the full spectral function probed here.
 
In the weak coupling limit $|k_Fa|\ll1$, the form of the T matrix
allows us to compute the leading corrections to
Eqs.~(\ref{eq:short},\ref{eq:shortrs2}), which are respectively given
by
\begin{align*}
    - \frac{2\sqrt{\pi} (1+i)
       k_F^3}{45\pi^2 a^2\sqrt{m_r^5}}  \ t^{5/2} , 
\ & \ \ - \frac{4\sqrt{\pi} (1-i)
       k_F^3}{315\pi^2 {R^*}^3 a \sqrt{m_r^7}}  \ t^{7/2} .
\end{align*}
These corrections yield the lowest orders at which the scattering
length enters the Ramsey response.

\begin{figure*}[th]
        \centering
        \includegraphics[width=0.74\textwidth]{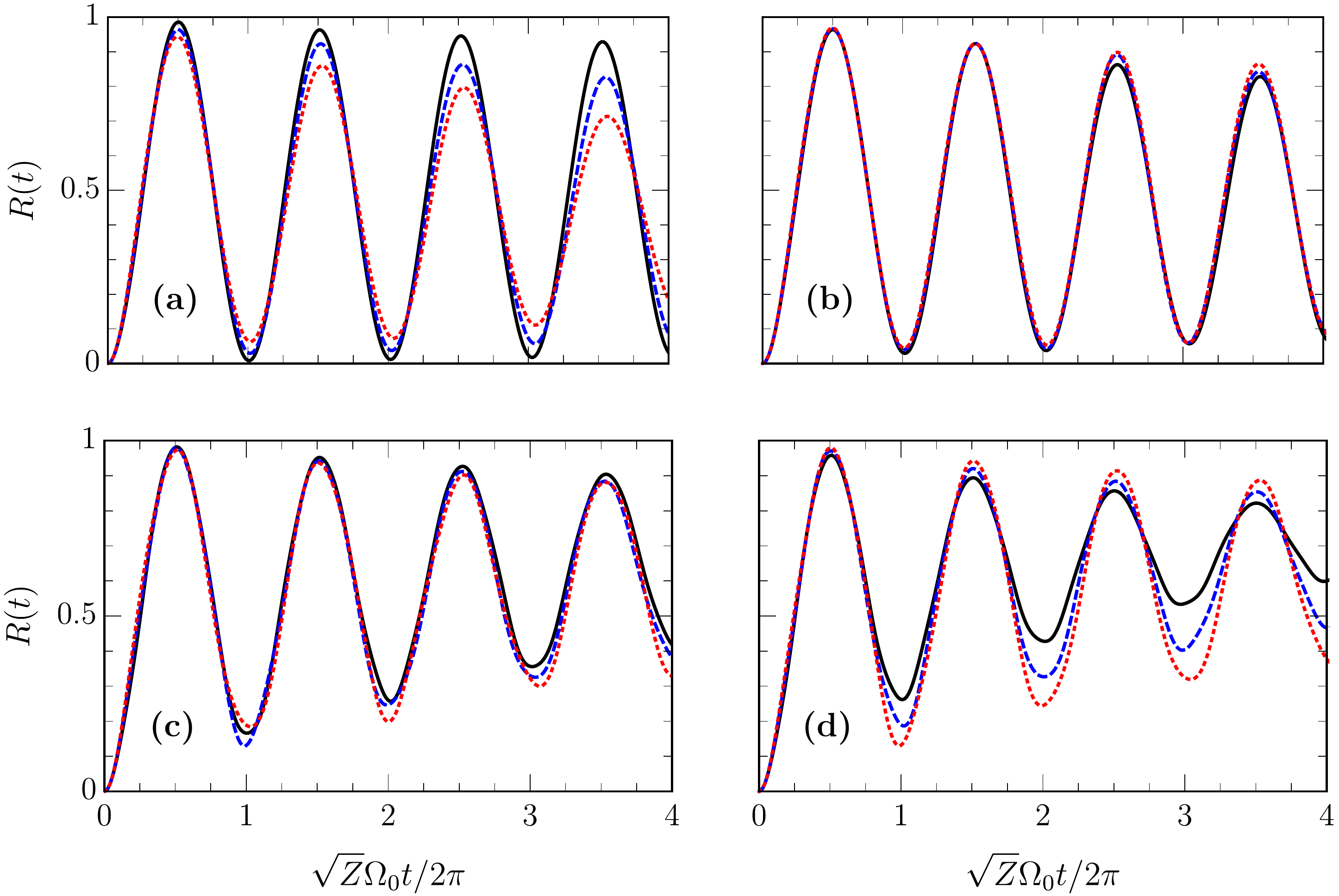}
        \caption{Occupation of the $\up$ impurity state, $R(t)$, as a
          function of time under a rf driving field of strength
          $\Omega_0$ that couples to the attractive $\up$ polaron
          (a,b) and the repulsive $\up$ polaron (c,d). (a) The
          response at unitarity and $R^*=0$ for different Rabi
          frequencies $\Omega_0 = 0.5 \ef$ (black solid), $0.75 \ef$
          (blue dashed), and $\ef$ (red dotted). (b) The response at
          fixed $\Omega_0 = 0.75 \ef$ for different
          $k_F R^* = 0, 0.5, 1$ with $1/k_Fa = 0, -0.277, -0.505$,
          (black solid, blue dashed, and red dotted), respectively.
          (c) Same Rabi frequencies as in (a), but with $k_F R^* = 1$
          and $1/k_Fa =1$.  (d) The response at fixed
          $\Omega_0 = 0.75 \ef$ and $1/k_Fa =1$ for different
          $k_F R^* = 0.1, 0.5, 1$, (black solid, blue dashed, and red
          dotted), respectively.  The conditions are arranged such
          that the residue of the $\up$ polaron is always
          $Z\simeq 0.784$.}
        \label{fig:rabi} 
\end{figure*}

Our results \eqref{eq:short} and \eqref{eq:shortrs2} are valid also
for a finite temperature $T$ provided that the time $t$ is shorter
than the characteristic time scale $1/T$ at which thermal effects
become relevant.  Likewise, for a finite impurity momentum $\p$, it is
clear from the form of the T matrix that there always exists a
frequency above which $\epsilon_{\p,\text{im}}$ is negligible and
therefore our results remain unchanged for
$t\ll 1/\epsilon_{\p,\text{im}}$. On the other hand our results are,
in general, sensitive to the preparation of the initial state.

\section{Rabi oscillations \label{sec:rabi}} 
Another important example of coherent impurity dynamics is the Rabi
oscillations between $\down$ and $\up$ impurity states that are driven
by a continuous rf field. The presence of the Fermi medium has an
observable effect on the oscillations when the spin-$\up$ state is
strongly interacting with the majority fermions.  For concreteness, we
assume that the impurity atom is initially in the non-interacting
$\down$ ground state, i.e.,
$\ket{\psi_R(t=0)}=\hat c_{\0\down}\ket{\text{FS}}$. Then, at times
$t\geq0$ the spin-$\down$ impurity is coupled to the interacting state
by the Rabi term in the Hamiltonian --- see Eq.~\eqref{eq:Haux}.  By
adjusting the rf detuning $\delta$ to match the attractive or
repulsive polaron energy, we can address either of these quasiparticle
branches.  Unlike in the perfect quench Ramsey response, Rabi
oscillations require us to take both impurity spin states explicitly
into account, and we thus employ the TBM with wave functions of the
form \eqref{eq:tbm} to describe the dynamics.  Specifically, we are
interested in the spin-$\up$ population
$N_\up = \bra{\psi_R(t)}\sum_\k \hat{c}_{\k\up}^\dag
\hat{c}_{\k\up}\ket{\psi_R(t)}$,
where the time-dependent wave function
$\ket{\psi_R(t)} = e^{-i\hat{H} t}\, \hat c_{\0\down}\ket{\text{FS}}$.

In Fig.~\ref{fig:rabi}, we show the relative occupation of the
spin-$\up$ impurity state, $R(t) = N_\up/(N_\up+N_\down)$, as a
function of time for Rabi frequencies typical in experiment. In panels
(a,b), the rf field addresses the attractive polaron, while (c,d) show
the results for the repulsive polaron. In all cases, regardless of the
interactions or the range parameter, we observe that the spin-$\up$
occupation displays a damped oscillatory behavior with period
$2\pi/(\sqrt{Z}\Omega_0)$. In other words, the angular frequency of
the Rabi oscillation in the presence of the Fermi medium is reduced by
a factor $\sqrt{Z}$ compared with that expected for a non-interacting
spin-$\up$ state.  This observation is consistent with the spin-$\up$
spectral function being dominated by the quasiparticle peaks, as in
Fig.~\ref{fig:spectra}. Specifically, if we assume that the spectrum
only contains the addressed quasiparticle, then one
obtains~\cite{Kohstall2012mac}
\begin{align}
R(t) &\simeq \sin^2(\sqrt{Z}\Omega_0 t/2) .
\end{align}
This reduction of the Rabi frequency has been used as a means to
experimentally access the polaron residue \cite{Kohstall2012mac}.

\begin{figure*}[th]
        \centering
        \includegraphics[width=0.73\textwidth]{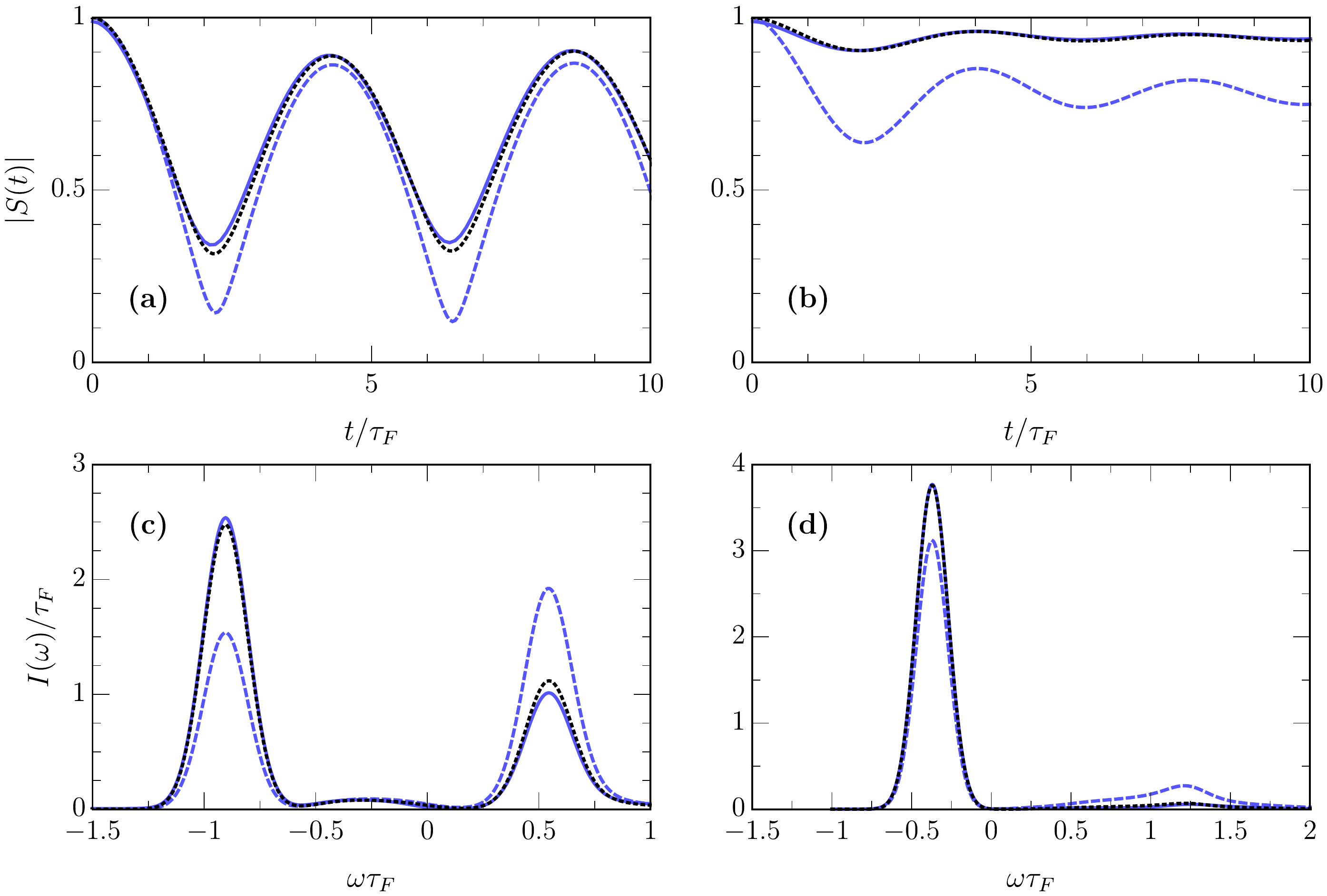}
        \caption{The Ramsey response $S(t)$ and corresponding
          convolved Fourier transform $I(\omega)$, with frequency
          broadening $\sigma = 0.1 \ef$. The interactions in the
          strongly interacting regime are (a,c) repulsive with
          $1/k_Fa=0.6$, and (b,d) attractive with $1/k_Fa=-0.5$. The
          range parameter is always taken to be $k_F R^*=1$. We show
          the result of a perfect quench to strong interactions
          (dashed), a state preparation with $t_\text{rf}=3\tau_F$ at
          $1/k_Fa_0=-2$ (solid), and the approximation
          Eq.~\eqref{eq:Sapprox} (dotted). In the second case,
            we include a wait time of $0.5 \tau_F$ at $1/k_Fa_0$ just
            after (before) the first (second) rf pulse to mimic the
            quantum state preparation in experiment.}
        \label{fig:prep} 
\end{figure*}

The manner of damping and the functional form of the Rabi oscillations
appears different in the four panels of Fig.~\ref{fig:rabi}. In (a,c)
we investigate the effect of changing the Rabi frequency $\Omega_0$ at
fixed interaction strength, whereas in (b,d) we change the interaction
strength while keeping $\Omega_0$ fixed. We see that when we address
the attractive polaron, the damping is sensitive to the bare Rabi
frequency, but quite insensitive to the precise interaction
parameters. The opposite appears to be the case for the repulsive
polaron, where the finite quasiparticle lifetime dominates the damping
and depends sensitively on the interaction parameters $1/k_Fa$ and
$k_F R^*$. This prediction is, in principle, straightforward to test
experimentally for sufficiently low temperatures. In practice, there
will also be damping due to thermal effects once $t > 1/T$.

\section{Quantum state preparation \label{sec:stateprep}}
For the Ramsey response $S(t)$, we have thus far considered the
situation of a perfect quench, where there is no effect of the medium
during the $\pi/2$ rf pulses, either because the pulses are infinitely
fast or because the impurity-medium interactions are switched off
during the pulses.  However, it is important to understand how such
`residual' interactions with the medium affect the impurity dynamics
since a perfect quench is challenging to achieve in
practice~\cite{Cetina2016}. Furthermore, one could in principle use
the residual interactions to tailor the initial state and engineer the
desired dynamical response.

To assess this effect in detail, we once again calculate the response
following the rf sequence in Fig.~\ref{fig:expt}, but this time we
consider the full Hamiltonian and wave function \eqref{eq:tbm} rather
than the decoupled spin-$\up$ versions. We address the attractive
polaron branch for interaction $1/k_Fa_0$ during the rf pulses by
setting the rf detuning $\delta = E_{\rm att}$, and then we extract
$S(t)$ as defined in Eq.~\eqref{eq:Ndiff}. We use the Rabi frequency
$\Omega_0/\ef = \pi/6$, which is typical in experiment.  In a previous
work~\cite{Cetina2016}, we employed such an approach for the case of a
heavy impurity and weak interactions, $|k_Fa_0| \ll1$, during the rf
pulses. Here, we consider stronger interactions on the attractive side
of resonance for the case of equal masses, as shown in
Fig.~\ref{fig:prep}. Compared to the result for the perfect quench, we
see that the residual attractive interactions clearly favour the
attractive polaron branch in the spectral function. This results in a
decrease in the amplitude of oscillations in the time domain and an
overall increase in the contrast $|S(t)|$ for the interactions
considered.  In principle, one could have a scenario where the
residual interactions \textit{increase} the amplitude of oscillations,
but this requires a larger $1/k_Fa$, %>1
where the TBM with a single particle-hole excitation becomes
increasingly inaccurate for the attractive branch.

Further insight can be gained by restricting ourselves to the
spin-$\up$ subspace and considering the approximate expression
\begin{align}
    S(t)\simeq e^{iE_0t} \bra{\psi_\text{att}}e^{-i\hat{H}_{\rm int}t}\ket{\psi_\text{att}},
    \label{eq:Sapprox}
\end{align}
where $\ket{\psi_\text{att}}$ is the attractive polaron state at
interaction parameter $1/k_Fa_0$ during the rf pulse. Referring to
Fig.~\ref{fig:prep}, we see that this approximation well reproduces
both the dynamical and spectral response obtained from the full
calculation.  Thus, we conclude that the main effect of the
interactions during the rf pulses is to \textit{adiabatically} prepare
attractive polaron states starting from the non-interacting wave
function. We expect this situation to hold provided the pulse duration
is longer than the Fermi time $\tau_F$.  In the limit
$\Omega \to \infty$, where the $\pi/2$ rf pulses become infinitely
fast, we should recover the perfect quench scenario from
Sec.~\ref{sec:response}.  In general, one could consider preparing
other initial, or reference, wave functions, which would shape the
final Ramsey response.

\begin{figure*}[th]
        \centering
        \includegraphics[width=\textwidth]{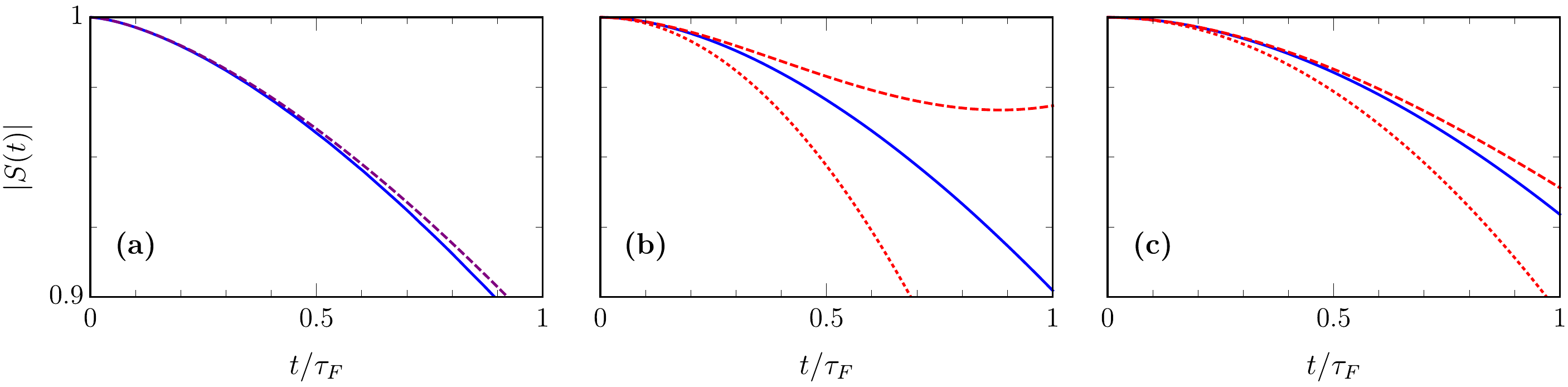}
        \caption{Short time behavior of the Ramsey contrast $|S(t)|$
          at unitarity for different values of the range parameter:
          (a) $k_FR^*=0$, (b) $k_FR^*=1/2$, and (c) $k_FR^*=1$. The
          result of the TBM where the wave functions are restricted to
          one particle-hole excitation as in Eq.~\eqref{eq:tbm} is
          shown as a solid line. In (a) this is compared with the
          short-time expansion for a broad resonance,
          Eq.~\eqref{eq:short} (dashed line). In (b,c) the short-time
          expansion at finite $R^*$ restricted to the ${\cal O}(t^2)$
          correction, Eq.~\eqref{eq:shortrs}, is shown as a dotted
          line, and the short-time expansion including the
          ${\cal O}(t^{5/2})$ correction, Eq.~\eqref{eq:shortrs2}, as
          a dashed line.  On this scale, the result of the TBM with
          two particle-hole excitations is indistinguishable from the
          solid line.}
        \label{fig:short} 
\end{figure*}

\begin{figure*}[th]
        \centering
        \includegraphics[width=\textwidth]{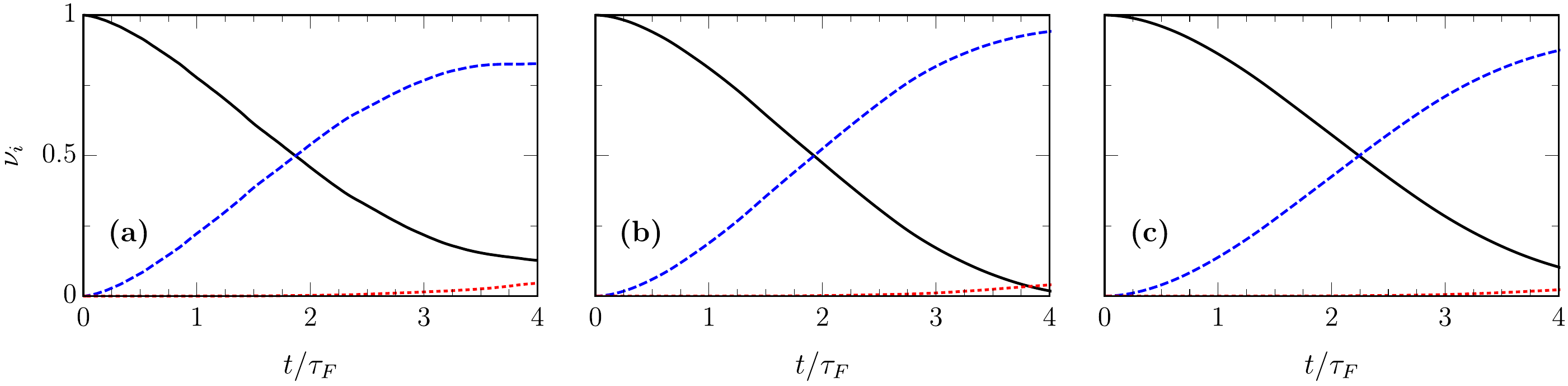}
        \caption{ Probability of $\psi(t)$ to be in states with 0
          (solid), 1 (dashed), and 2 (dotted) holes excited from the
          Fermi sea. The results are shown at unitarity for (a)
          $k_FR^*=0$, (b) $k_FR^*=1/2$, and (c) $k_FR^*=1$.}
        \label{fig:nu} 
\end{figure*}

\section{Multiple particle-hole excitations}
\label{sec:2ph}
We now discuss the role played by multiple particle-hole excitations
in the impurity dynamics, quantified here by the Ramsey contrast for
the perfect quench. Going beyond the single particle-hole
approximation in the TBM is in general a complicated problem, as the
size of the truncated subspace grows exponentially with the number of
excitations of the Fermi sea.  Thus, in this section we focus on a
simpler problem than what has been described so far, namely that of a
static (infinitely heavy) impurity particle.  In this case, the
angular degrees of freedom can be integrated out, allowing us to
extend the wave function \eqref{eq:tbm} to two particle-hole
excitations (see Appendix \ref{app:2ph} for the mathematical details).

The static impurity problem may be solved exactly since it reduces to
the problem of a single particle in the presence of a fixed potential.
At the same time, we expect the static impurity to constitute a worst
case scenario for the TBM since it features the orthogonality
catastrophe~\cite{Nozieres1969sit}, where there is no well-defined
quasiparticle (i.e., the residue $Z=0$) and one has an infinite number
of low-energy excitations.  We previously compared the exact solution
with the TBM for one particle-hole excitation and found excellent
agreement for short times up to order $10 \tau_F$ near
unitarity~\cite{Cetina2016}.  Here, we analyse the structure of the
wave function and estimate the timescale at which multiple
particle-hole excitations appear for different range parameters
$k_F R^*$.

Consider first the Ramsey response at times $t\lesssim \tau_F$. In
this case, the analytic expressions for $|S(t)|$ at short times ---
Eq.~\eqref{eq:short} for $R^*=0$ and Eq.~\eqref{eq:shortrs2} for
$R^*>0$ --- were derived from the observation that the short-time
dynamics is governed by large frequencies and thus two-body
physics. Since the wave function with one particle-hole excitation,
Eq.~\eqref{eq:tbm}, explicitly includes the processes constituting the
two-body scattering $T$ matrix, we expect the short-time dynamics to
be well captured by this wave function. In Fig.~\ref{fig:short} we
show the Ramsey contrast at unitarity for $k_FR^*=0$, $1/2$, and 1,
and indeed we observe that both the initial $t^{3/2}$ decrease of the
contrast for $k_FR^*=0$, and the $t^2$ decrease plus $t^{5/2}$
correction for finite $k_FR^*$ are well captured by the TBM. At times
up to $\tau_F$, we furthermore find perfect agreement between the
results of diagonalizing the wave functions with one and two
particle-hole excitations, as the results are identical within our
numerical error (which, for $t\lesssim 4\tau_F$, we estimate to be
less than $0.01\%$ in the one particle-hole TBM and less than $1\%$
for $|S(t)|$ in the two particle-hole TBM).

At times exceeding $\tau_F$, eventually multiple particle-hole
excitations become important. In order to quantify the contribution
from the different terms in the variational wave function, we project
$\psi(t)$ including two particle-hole excitations (App.~\ref{app:2ph})
onto basis states with a fixed number --- 0, 1, or 2 --- of holes
excited from the Fermi sea:
\begin{align*}
\nu_0=& |\!\bra{\rm{FS}}\hat c_{\0\up}|\psi(t)\rangle|^2, \nn \\
\nu_1=&
\sum_\q|\!\bra{\rm{FS}}\hat f^\dagger_\q\hat d_\q|\psi(t)\rangle|^2+
\sum_{\k,\q}|\!\bra{\rm{FS}}\hat f^\dagger_\q\hat f_\k\hat c_{\q-\k \up}|\psi(t)\rangle|^2,
\nn \\
\nu_2=&\frac12\sum_{\k,\q_1,\q_2}|\!\bra{\rm{FS}}\hat
        f^\dagger_{\q_1}\hat f^\dagger_{\q_2}\hat f_\k\hat
        d_{\q_1+\q_2-\k}|\psi(t)\rangle|^2 \nn \\
& \hspace{-4mm}
+ \frac14 \sum_{\k_1,\k_2,\q_1,\q_2}|\!\bra{\rm{FS}}\hat
  f^\dagger_{\q_1} \hat f^\dagger_{\q_2}\hat f_{\k_1}\hat f_{\k_2}\hat
  c_{\q_1+\q_2-\k_1-\k_2 \up}|\psi(t)\rangle|^2. \nn 
\end{align*}
As previously, $\k,\,\k_1,\,\k_2$ denote particles above the Fermi sea
and $\q,\,\q_1,\,\q_2$ holes.  Note that we have $\nu_0=|S(t)|^2$ and
$\nu_0+\nu_1+\nu_2=1$. In Fig.~\ref{fig:nu} we display these
quantities at unitarity for various values of the resonance range. We
clearly see that two particle-hole excitations remain insignificant
even at several times the Fermi time. Furthermore, this result is
independent of the resonance range, and thus it is insensitive to the
precise power-law behavior at short times.

Since the static impurity is a worst case scenario, we expect the TBM
for wave functions \eqref{eq:tbm} to accurately describe the
short-time dynamics also for a \emph{mobile} impurity.

\section{Outlook}
\label{sec:conc}

In this work, we have shown how the truncated basis method can be used
to determine the coherent quantum evolution of an impurity for a
variety of scenarios, including the dynamical response to a suddenly
introduced impurity, and the Rabi oscillations between a polaron
quasiparticle and a non-interacting impurity state. We have
furthermore explored the connection between impurity dynamics and the
spectral function, as well as the possibility of preparing different
quantum many-body states.  We emphasize that the method is quite
general, and allows one to investigate more complicated dynamical
protocols. For instance, a spin-echo sequence can be efficiently
modelled as a series of time-evolution operators, each of which are
evaluated within the truncated basis.

A key result is the exact short-time Ramsey response, i.e., for
$t\lesssim \tau_F$. Surprisingly, this was found to depend
non-analytically on time, which is a direct consequence of the need to
renormalize the short-range interactions. Additionally, the response
to leading order does not depend on the scattering length, although
the regime of validity of the short-time expansion does. We also note
that our short-time expansion is not affected by temperature in a
degenerate gas, since the time-scale associated with thermal
excitations is longer than the Fermi time.  Therefore, our predictions
can be tested in current precision experiments on ultracold atomic
gases out of equilibrium \cite{Cetina2016}.

In the long-time limit $t\gg \tau_F$ (but still $t\lesssim 1/T$), we
have argued that if the ground state of the interacting impurity is a
well-defined attractive polaron, the Ramsey response will be dominated
by the corresponding single peak in the spectral function. In turn,
this limit will be well captured by the TBM, as it describes both the
energy and residue of the attractive polaron very well
\cite{combescot2009}. While the TBM thus captures both the short- and
the long-time coherent impurity dynamics well, the evolution at
intermediate times presents an outstanding challenge to theories of
strongly correlated quantum matter.

For a static impurity, we have shown that multiple particle-hole
excitations only become prominent in the Ramsey response at time
scales significantly exceeding the Fermi time.  Thus, while we expect
a power-law decay of the contrast $|S(t)|$ due to the orthogonality
catastrophe, this is driven by low-energy excitations which are only
relevant at long times. In the opposite limit of a light impurity,
the Ramsey response will likely develop exotic few-body
correlations. In this case, for sufficiently large mass ratio, the
system is predicted to feature universal trimers
\cite{Kartavtsev2007let} with associated resonant few-body
interactions \cite{Levinsen2009ads}, or even Efimov trimers
\cite{Efimov1973elo} and tetramers \cite{Castin2010fbe}. The
theoretical description of such systems would thus require one to go
to a truncated Hilbert space featuring multiple excitations of the
medium.

An open question is how to extend the present work to non-zero
temperature, where one requires a thermal average over all initial
states.  In particular, it would be interesting to understand how the
impurity dynamics evolves from the quantum short-time regime to the
thermal long-time limit.  When the interactions are weak
$|k_F a| < 1$, the long-time decoherence due to thermal fluctuations
is well described using a Fermi liquid calculation for quasiparticle
scattering~\cite{Cetina2015doi}.  A major simplification is to
approximate the impurity as being effectively fixed by the thermal
excitations of the medium, in which case the Ramsey response may be
obtained exactly using a functional determinant approach
\cite{Cetina2016}.  However, such an approximation is only expected to
be reasonable when $\ef m/m_{\rm im} < T < \ef$, i.e., for
sufficiently heavy impurities and sufficiently high
temperatures~\cite{rosch1995hpi}.

Finally, the framework developed here is not limited to a
single-component fermionic medium. As an example, we have recently
calculated the spectral response to an rf pulse for an impurity in a
Bose-Einstein condensate: Here we obtained a very good agreement
between the TBM, containing up to two Bogoliubov excitations of the
condensate, and the experimental measurements
\cite{Jorgensen2016}. Likewise, the TBM could be applied to the
problem of an impurity in a two-component Fermi gas across the BCS-BEC
crossover~\cite{Nishida2015,Wei2015}. A particularly interesting
future application would be to the investigation of three-body
correlations in a Bose-Einstein condensate that is suddenly quenched
to unitarity, as in experiment~\cite{Makotyn:2014fk}.

%%%%%%%%%%%%%%%%%%%%%%%%%%%%%%%%%%%%%%%%%%%%%%%%%
\acknowledgments

We gratefully acknowledge fruitful discussions with Masud Haque,
Richard Schmidt, Georg Bruun, Marko Cetina, Jook Walraven, and Rudi
Grimm.

%%%%%%%%%%%%%%%%%%%%%%%%%%%%%%%%%%%%%%%%%%%%%%%%%

\appendix

\begin{widetext}

\section{Variational wave functions with two particle-hole exciations}
\label{app:2ph}

In this appendix, we present the variational equations for a wave function with two particle-hole excitations:
\begin{align}
\ket{\psi} 
= & \left[ \alpha_{0} \hat c^\dag_{\mathbf{0}\up} 
 + \sum_\vect{q}  \alpha_\vect{q} \hat d^\dag_\vect{q} \hat f_{\vect{q}}   +  \sum_{\vect{k}, \vect{q}}
\alpha_{\vect{k}\vect{q}}\hat c^\dag_{\vect{q}- \vect{k}\up}
\hat f^\dag_{\vect{k}} \hat f_{\vect{q}}
 + \sum_\vect{\k,\q_1,\q_2}  \alpha_{\k\q_1\q_2} \hat d^\dag_{\q_1+\q_2-\k} \hat f^\dagger_\k\hat f_{\q_1}\hat f_{\q_2}   
 \right.
 \nn \\ & \left. +  
 \sum_{\k_1,\k_2,\q_1,\q_2}
\alpha_{\k_1\k_2\q_1\q_2}\hat c^\dag_{\q_1+\q_2-\k_1-\k_2\up}
\hat f^\dag_{\k_1}\hat f^\dag_{\k_2} \hat f_{\q_1} \hat f_{\q_2}
\right] \left| \rm{FS}\right>,
\end{align}
which provides a natural extension of the wave function \eqref{eq:tbm}. The variational equations then become
\begin{align}
(E-E_0)\alpha_0 & =g\sum_\q\alpha_\q \nn \\
(E-E_0)\alpha_\q & = (\epsilon_{\q,\text{M}}-\epsilon_\q+\nu)\alpha_\q+g\alpha_0+q\sum_\k \alpha_{\k\q} \nn \\
(E-E_0)\alpha_{\k\q} & =(\epsilon_{\q-\k,\text{im}}+\ek-\eq)\alpha_{\k\q}
+g\alpha_\q+g\sum_{\q'}\alpha_{\k\q\q'} \nn\\
(E-E_0)\alpha_{\k\q_1\q_2} & =(\epsilon_{\q_1+\q_2-\k,\text{M}}+\ek-\epsilon_{\q_1}-\epsilon_{\q_2}-\nu)\alpha_{\k\q_1\q_2}+g\alpha_{\k\q_1}-g\alpha_{\k\q_2} +g\sum_{\k'}\alpha_{\k'\k\q_1\q_2} \nn \\
(E-E_0)\alpha_{\k_1\k_2\q_1\q_2} & =(\epsilon_{\q_1+\q_2-\k_1-\k_2,\text{im}}+
\epsilon_{\k_1}+\epsilon_{\k_2}-\epsilon_{\q_1}-\epsilon_{\q_2})\alpha_{\k_1\k_2\q_1\q_2}+g(\alpha_{\k_2\q_1\q_2}-\alpha_{\k_1\q_1\q_2}),
\end{align}
where we have used $\alpha_{\k\q_1\q_2}=-\alpha_{\k\q_2\q_1}$ and $\alpha_{\k_1\k_2\q_1\q_2}=-\alpha_{\k_2\k_1\q_1\q_2}=-\alpha_{\k_1\k_2\q_2\q_1}$. For a static impurity, we have $\epsilon_{\k,\text{im}}=\epsilon_{\k,\text{M}}=0$, and thus the equations become independent of the angles between vectors.
\end{widetext}

\bibliography{ultracold,library,fastdyn,polaron_refs}

\begin{thebibliography}{51}%
\makeatletter
\providecommand \@ifxundefined [1]{%
 \@ifx{#1\undefined}
}%
\providecommand \@ifnum [1]{%
 \ifnum #1\expandafter \@firstoftwo
 \else \expandafter \@secondoftwo
 \fi
}%
\providecommand \@ifx [1]{%
 \ifx #1\expandafter \@firstoftwo
 \else \expandafter \@secondoftwo
 \fi
}%
\providecommand \natexlab [1]{#1}%
\providecommand \enquote  [1]{``#1''}%
\providecommand \bibnamefont  [1]{#1}%
\providecommand \bibfnamefont [1]{#1}%
\providecommand \citenamefont [1]{#1}%
\providecommand \href@noop [0]{\@secondoftwo}%
\providecommand \href [0]{\begingroup \@sanitize@url \@href}%
\providecommand \@href[1]{\@@startlink{#1}\@@href}%
\providecommand \@@href[1]{\endgroup#1\@@endlink}%
\providecommand \@sanitize@url [0]{\catcode `\\12\catcode `\$12\catcode
  `\&12\catcode `\#12\catcode `\^12\catcode `\_12\catcode `\%12\relax}%
\providecommand \@@startlink[1]{}%
\providecommand \@@endlink[0]{}%
\providecommand \url  [0]{\begingroup\@sanitize@url \@url }%
\providecommand \@url [1]{\endgroup\@href {#1}{\urlprefix }}%
\providecommand \urlprefix  [0]{URL }%
\providecommand \Eprint [0]{\href }%
\providecommand \doibase [0]{http://dx.doi.org/}%
\providecommand \selectlanguage [0]{\@gobble}%
\providecommand \bibinfo  [0]{\@secondoftwo}%
\providecommand \bibfield  [0]{\@secondoftwo}%
\providecommand \translation [1]{[#1]}%
\providecommand \BibitemOpen [0]{}%
\providecommand \bibitemStop [0]{}%
\providecommand \bibitemNoStop [0]{.\EOS\space}%
\providecommand \EOS [0]{\spacefactor3000\relax}%
\providecommand \BibitemShut  [1]{\csname bibitem#1\endcsname}%
\let\auto@bib@innerbib\@empty
%</preamble>
\bibitem [{\citenamefont {Bloch}\ \emph {et~al.}(2008)\citenamefont {Bloch},
  \citenamefont {Dalibard},\ and\ \citenamefont {Zwerger}}]{Bloch2008}%
  \BibitemOpen
  \bibfield  {author} {\bibinfo {author} {\bibfnamefont {I.}~\bibnamefont
  {Bloch}}, \bibinfo {author} {\bibfnamefont {J.}~\bibnamefont {Dalibard}}, \
  and\ \bibinfo {author} {\bibfnamefont {W.}~\bibnamefont {Zwerger}},\ }\href
  {\doibase 10.1103/RevModPhys.80.885} {\bibfield  {journal} {\bibinfo
  {journal} {Rev. Mod. Phys.}\ }\textbf {\bibinfo {volume} {80}},\ \bibinfo
  {pages} {885} (\bibinfo {year} {2008})}\BibitemShut {NoStop}%
\bibitem [{\citenamefont {Barankov}\ \emph {et~al.}(2004)\citenamefont
  {Barankov}, \citenamefont {Levitov},\ and\ \citenamefont
  {Spivak}}]{Barankov2004}%
  \BibitemOpen
  \bibfield  {author} {\bibinfo {author} {\bibfnamefont {R.~A.}\ \bibnamefont
  {Barankov}}, \bibinfo {author} {\bibfnamefont {L.~S.}\ \bibnamefont
  {Levitov}}, \ and\ \bibinfo {author} {\bibfnamefont {B.~Z.}\ \bibnamefont
  {Spivak}},\ }\href {\doibase 10.1103/PhysRevLett.93.160401} {\bibfield
  {journal} {\bibinfo  {journal} {Phys. Rev. Lett.}\ }\textbf {\bibinfo
  {volume} {93}},\ \bibinfo {pages} {160401} (\bibinfo {year}
  {2004})}\BibitemShut {NoStop}%
\bibitem [{\citenamefont {Andreev}\ \emph {et~al.}(2004)\citenamefont
  {Andreev}, \citenamefont {Gurarie},\ and\ \citenamefont
  {Radzihovsky}}]{Andreev2004}%
  \BibitemOpen
  \bibfield  {author} {\bibinfo {author} {\bibfnamefont {A.~V.}\ \bibnamefont
  {Andreev}}, \bibinfo {author} {\bibfnamefont {V.}~\bibnamefont {Gurarie}}, \
  and\ \bibinfo {author} {\bibfnamefont {L.}~\bibnamefont {Radzihovsky}},\
  }\href {\doibase 10.1103/PhysRevLett.93.130402} {\bibfield  {journal}
  {\bibinfo  {journal} {Phys. Rev. Lett.}\ }\textbf {\bibinfo {volume} {93}},\
  \bibinfo {pages} {130402} (\bibinfo {year} {2004})}\BibitemShut {NoStop}%
\bibitem [{\citenamefont {Szyma\ifmmode~\acute{n}\else \'{n}\fi{}ska}\ \emph
  {et~al.}(2005)\citenamefont {Szyma\ifmmode~\acute{n}\else \'{n}\fi{}ska},
  \citenamefont {Simons},\ and\ \citenamefont {Burnett}}]{Szymanska2005}%
  \BibitemOpen
  \bibfield  {author} {\bibinfo {author} {\bibfnamefont {M.~H.}\ \bibnamefont
  {Szyma\ifmmode~\acute{n}\else \'{n}\fi{}ska}}, \bibinfo {author}
  {\bibfnamefont {B.~D.}\ \bibnamefont {Simons}}, \ and\ \bibinfo {author}
  {\bibfnamefont {K.}~\bibnamefont {Burnett}},\ }\href {\doibase
  10.1103/PhysRevLett.94.170402} {\bibfield  {journal} {\bibinfo  {journal}
  {Phys. Rev. Lett.}\ }\textbf {\bibinfo {volume} {94}},\ \bibinfo {pages}
  {170402} (\bibinfo {year} {2005})}\BibitemShut {NoStop}%
\bibitem [{\citenamefont {Zwierlein}\ \emph
  {et~al.}(2006{\natexlab{a}})\citenamefont {Zwierlein}, \citenamefont
  {Schirotzek}, \citenamefont {Schunck},\ and\ \citenamefont
  {Ketterle}}]{Zwierlein2006}%
  \BibitemOpen
  \bibfield  {author} {\bibinfo {author} {\bibfnamefont {M.~W.}\ \bibnamefont
  {Zwierlein}}, \bibinfo {author} {\bibfnamefont {A.}~\bibnamefont
  {Schirotzek}}, \bibinfo {author} {\bibfnamefont {C.~H.}\ \bibnamefont
  {Schunck}}, \ and\ \bibinfo {author} {\bibfnamefont {W.}~\bibnamefont
  {Ketterle}},\ }\href {\doibase 10.1126/science.1122318} {\bibfield  {journal}
  {\bibinfo  {journal} {Science}\ }\textbf {\bibinfo {volume} {311}},\ \bibinfo
  {pages} {492} (\bibinfo {year} {2006}{\natexlab{a}})}\BibitemShut {NoStop}%
\bibitem [{\citenamefont {Zwierlein}\ \emph
  {et~al.}(2006{\natexlab{b}})\citenamefont {Zwierlein}, \citenamefont
  {Schirotzek}, \citenamefont {Schunck},\ and\ \citenamefont
  {Ketterle}}]{Zwierlein2006doo}%
  \BibitemOpen
  \bibfield  {author} {\bibinfo {author} {\bibfnamefont {M.~W.}\ \bibnamefont
  {Zwierlein}}, \bibinfo {author} {\bibfnamefont {A.}~\bibnamefont
  {Schirotzek}}, \bibinfo {author} {\bibfnamefont {C.~H.}\ \bibnamefont
  {Schunck}}, \ and\ \bibinfo {author} {\bibfnamefont {W.}~\bibnamefont
  {Ketterle}},\ }\href {\doibase 10.1038/nature04936} {\bibfield  {journal}
  {\bibinfo  {journal} {Nature (London)}\ }\textbf {\bibinfo {volume} {442}},\
  \bibinfo {pages} {54} (\bibinfo {year} {2006}{\natexlab{b}})}\BibitemShut
  {NoStop}%
\bibitem [{\citenamefont {Partridge}\ \emph {et~al.}(2006)\citenamefont
  {Partridge}, \citenamefont {Li}, \citenamefont {Kamar}, \citenamefont
  {Liao},\ and\ \citenamefont {Hulet}}]{Partridge2006pap}%
  \BibitemOpen
  \bibfield  {author} {\bibinfo {author} {\bibfnamefont {G.~B.}\ \bibnamefont
  {Partridge}}, \bibinfo {author} {\bibfnamefont {W.}~\bibnamefont {Li}},
  \bibinfo {author} {\bibfnamefont {R.~I.}\ \bibnamefont {Kamar}}, \bibinfo
  {author} {\bibfnamefont {Y.}~\bibnamefont {Liao}}, \ and\ \bibinfo {author}
  {\bibfnamefont {R.~G.}\ \bibnamefont {Hulet}},\ }\href {\doibase
  10.1126/science.1122876} {\bibfield  {journal} {\bibinfo  {journal}
  {Science}\ }\textbf {\bibinfo {volume} {311}},\ \bibinfo {pages} {503}
  (\bibinfo {year} {2006})}\BibitemShut {NoStop}%
\bibitem [{\citenamefont {Schirotzek}\ \emph {et~al.}(2009)\citenamefont
  {Schirotzek}, \citenamefont {Wu}, \citenamefont {Sommer},\ and\ \citenamefont
  {Zwierlein}}]{Schirotzek2009oof}%
  \BibitemOpen
  \bibfield  {author} {\bibinfo {author} {\bibfnamefont {A.}~\bibnamefont
  {Schirotzek}}, \bibinfo {author} {\bibfnamefont {C.-H.}\ \bibnamefont {Wu}},
  \bibinfo {author} {\bibfnamefont {A.}~\bibnamefont {Sommer}}, \ and\ \bibinfo
  {author} {\bibfnamefont {M.~W.}\ \bibnamefont {Zwierlein}},\ }\href {\doibase
  10.1103/PhysRevLett.102.230402} {\bibfield  {journal} {\bibinfo  {journal}
  {Phys. Rev. Lett.}\ }\textbf {\bibinfo {volume} {102}},\ \bibinfo {pages}
  {230402} (\bibinfo {year} {2009})}\BibitemShut {NoStop}%
\bibitem [{\citenamefont {Nascimb\`ene}\ \emph {et~al.}(2009)\citenamefont
  {Nascimb\`ene}, \citenamefont {Navon}, \citenamefont {Jiang}, \citenamefont
  {Tarruell}, \citenamefont {Teichmann}, \citenamefont {McKeever},
  \citenamefont {Chevy},\ and\ \citenamefont {Salomon}}]{Nascimbene2009}%
  \BibitemOpen
  \bibfield  {author} {\bibinfo {author} {\bibfnamefont {S.}~\bibnamefont
  {Nascimb\`ene}}, \bibinfo {author} {\bibfnamefont {N.}~\bibnamefont {Navon}},
  \bibinfo {author} {\bibfnamefont {K.~J.}\ \bibnamefont {Jiang}}, \bibinfo
  {author} {\bibfnamefont {L.}~\bibnamefont {Tarruell}}, \bibinfo {author}
  {\bibfnamefont {M.}~\bibnamefont {Teichmann}}, \bibinfo {author}
  {\bibfnamefont {J.}~\bibnamefont {McKeever}}, \bibinfo {author}
  {\bibfnamefont {F.}~\bibnamefont {Chevy}}, \ and\ \bibinfo {author}
  {\bibfnamefont {C.}~\bibnamefont {Salomon}},\ }\href {\doibase
  10.1103/PhysRevLett.103.170402} {\bibfield  {journal} {\bibinfo  {journal}
  {Phys. Rev. Lett.}\ }\textbf {\bibinfo {volume} {103}},\ \bibinfo {pages}
  {170402} (\bibinfo {year} {2009})}\BibitemShut {NoStop}%
\bibitem [{\citenamefont {Liao}\ \emph {et~al.}(2010)\citenamefont {Liao},
  \citenamefont {Rittner}, \citenamefont {Paprotta}, \citenamefont {Li},
  \citenamefont {Partridge}, \citenamefont {Hulet}, \citenamefont {Baur},\ and\
  \citenamefont {Mueller}}]{Liao:2010bh}%
  \BibitemOpen
  \bibfield  {author} {\bibinfo {author} {\bibfnamefont {Y.-a.}\ \bibnamefont
  {Liao}}, \bibinfo {author} {\bibfnamefont {A.~S.~C.}\ \bibnamefont
  {Rittner}}, \bibinfo {author} {\bibfnamefont {T.}~\bibnamefont {Paprotta}},
  \bibinfo {author} {\bibfnamefont {W.}~\bibnamefont {Li}}, \bibinfo {author}
  {\bibfnamefont {G.~B.}\ \bibnamefont {Partridge}}, \bibinfo {author}
  {\bibfnamefont {R.~G.}\ \bibnamefont {Hulet}}, \bibinfo {author}
  {\bibfnamefont {S.~K.}\ \bibnamefont {Baur}}, \ and\ \bibinfo {author}
  {\bibfnamefont {E.~J.}\ \bibnamefont {Mueller}},\ }\href
  {http://dx.doi.org/10.1038/nature09393} {\bibfield  {journal} {\bibinfo
  {journal} {Nature}\ }\textbf {\bibinfo {volume} {467}},\ \bibinfo {pages}
  {567} (\bibinfo {year} {2010})}\BibitemShut {NoStop}%
\bibitem [{\citenamefont {Kohstall}\ \emph {et~al.}(2012)\citenamefont
  {Kohstall}, \citenamefont {Zaccanti}, \citenamefont {Jag}, \citenamefont
  {Trenkwalder}, \citenamefont {Massignan}, \citenamefont {Bruun},
  \citenamefont {Schreck},\ and\ \citenamefont {Grimm}}]{Kohstall2012mac}%
  \BibitemOpen
  \bibfield  {author} {\bibinfo {author} {\bibfnamefont {C.}~\bibnamefont
  {Kohstall}}, \bibinfo {author} {\bibfnamefont {M.}~\bibnamefont {Zaccanti}},
  \bibinfo {author} {\bibfnamefont {M.}~\bibnamefont {Jag}}, \bibinfo {author}
  {\bibfnamefont {A.}~\bibnamefont {Trenkwalder}}, \bibinfo {author}
  {\bibfnamefont {P.}~\bibnamefont {Massignan}}, \bibinfo {author}
  {\bibfnamefont {G.~M.}\ \bibnamefont {Bruun}}, \bibinfo {author}
  {\bibfnamefont {F.}~\bibnamefont {Schreck}}, \ and\ \bibinfo {author}
  {\bibfnamefont {R.}~\bibnamefont {Grimm}},\ }\href {\doibase
  10.1038/nature11065} {\bibfield  {journal} {\bibinfo  {journal} {Nature
  (London)}\ }\textbf {\bibinfo {volume} {485}},\ \bibinfo {pages} {615}
  (\bibinfo {year} {2012})}\BibitemShut {NoStop}%
\bibitem [{\citenamefont {Koschorreck}\ \emph {et~al.}(2012)\citenamefont
  {Koschorreck}, \citenamefont {Pertot}, \citenamefont {Vogt}, \citenamefont
  {Fr\"{o}lich}, \citenamefont {Feld},\ and\ \citenamefont
  {K\"{o}hl}}]{Koschorreck2012aar}%
  \BibitemOpen
  \bibfield  {author} {\bibinfo {author} {\bibfnamefont {M.}~\bibnamefont
  {Koschorreck}}, \bibinfo {author} {\bibfnamefont {D.}~\bibnamefont {Pertot}},
  \bibinfo {author} {\bibfnamefont {E.}~\bibnamefont {Vogt}}, \bibinfo {author}
  {\bibfnamefont {B.}~\bibnamefont {Fr\"{o}lich}}, \bibinfo {author}
  {\bibfnamefont {M.}~\bibnamefont {Feld}}, \ and\ \bibinfo {author}
  {\bibfnamefont {M.}~\bibnamefont {K\"{o}hl}},\ }\href {\doibase
  10.1038/nature11151} {\bibfield  {journal} {\bibinfo  {journal} {Nature
  (London)}\ }\textbf {\bibinfo {volume} {485}},\ \bibinfo {pages} {619}
  (\bibinfo {year} {2012})}\BibitemShut {NoStop}%
\bibitem [{\citenamefont {{Mitra}}\ \emph {et~al.}(2016)\citenamefont
  {{Mitra}}, \citenamefont {{Brown}}, \citenamefont {{Schau{\ss}}},
  \citenamefont {{Kondov}},\ and\ \citenamefont {{Bakr}}}]{Bakr2016}%
  \BibitemOpen
  \bibfield  {author} {\bibinfo {author} {\bibfnamefont {D.}~\bibnamefont
  {{Mitra}}}, \bibinfo {author} {\bibfnamefont {P.~T.}\ \bibnamefont
  {{Brown}}}, \bibinfo {author} {\bibfnamefont {P.}~\bibnamefont
  {{Schau{\ss}}}}, \bibinfo {author} {\bibfnamefont {S.~S.}\ \bibnamefont
  {{Kondov}}}, \ and\ \bibinfo {author} {\bibfnamefont {W.~S.}\ \bibnamefont
  {{Bakr}}},\ }\href@noop {} {\bibfield  {journal} {\bibinfo  {journal} {ArXiv
  e-prints}\ } (\bibinfo {year} {2016})},\ \Eprint
  {http://arxiv.org/abs/1604.01479} {arXiv:1604.01479 [cond-mat.quant-gas]}
  \BibitemShut {NoStop}%
\bibitem [{\citenamefont {Nozi\`eres}\ and\ \citenamefont
  {de~Dominicis}(1969)}]{Nozieres1969sit}%
  \BibitemOpen
  \bibfield  {author} {\bibinfo {author} {\bibfnamefont {P.}~\bibnamefont
  {Nozi\`eres}}\ and\ \bibinfo {author} {\bibfnamefont {C.~T.}\ \bibnamefont
  {de~Dominicis}},\ }\href {\doibase 10.1103/PhysRev.178.1097} {\bibfield
  {journal} {\bibinfo  {journal} {Phys. Rev.}\ }\textbf {\bibinfo {volume}
  {178}},\ \bibinfo {pages} {1097} (\bibinfo {year} {1969})}\BibitemShut
  {NoStop}%
\bibitem [{\citenamefont {Cetina}\ \emph {et~al.}()\citenamefont {Cetina},
  \citenamefont {Jag}, \citenamefont {Lous}, \citenamefont {Fritsche},
  \citenamefont {Walraven}, \citenamefont {Grimm}, \citenamefont {Levinsen},
  \citenamefont {Parish}, \citenamefont {Schmidt}, \citenamefont {Knap},\ and\
  \citenamefont {Demler}}]{Cetina2016}%
  \BibitemOpen
  \bibfield  {author} {\bibinfo {author} {\bibfnamefont {M.}~\bibnamefont
  {Cetina}}, \bibinfo {author} {\bibfnamefont {M.}~\bibnamefont {Jag}},
  \bibinfo {author} {\bibfnamefont {R.~S.}\ \bibnamefont {Lous}}, \bibinfo
  {author} {\bibfnamefont {I.}~\bibnamefont {Fritsche}}, \bibinfo {author}
  {\bibfnamefont {J.~T.~M.}\ \bibnamefont {Walraven}}, \bibinfo {author}
  {\bibfnamefont {R.}~\bibnamefont {Grimm}}, \bibinfo {author} {\bibfnamefont
  {J.}~\bibnamefont {Levinsen}}, \bibinfo {author} {\bibfnamefont {M.~M.}\
  \bibnamefont {Parish}}, \bibinfo {author} {\bibfnamefont {R.}~\bibnamefont
  {Schmidt}}, \bibinfo {author} {\bibfnamefont {M.}~\bibnamefont {Knap}}, \
  and\ \bibinfo {author} {\bibfnamefont {E.}~\bibnamefont {Demler}},\
  }\href@noop {} {\enquote {\bibinfo {title} {Ultrafast many-body
  interferometry of impurities coupled to a {Fermi} sea},}\ }\bibinfo {note}
  {ArXiv:1604.07423}\BibitemShut {NoStop}%
\bibitem [{\citenamefont {Goold}\ \emph {et~al.}(2011)\citenamefont {Goold},
  \citenamefont {Fogarty}, \citenamefont {Lo~Gullo}, \citenamefont
  {Paternostro},\ and\ \citenamefont {Busch}}]{Goold2011oca}%
  \BibitemOpen
  \bibfield  {author} {\bibinfo {author} {\bibfnamefont {J.}~\bibnamefont
  {Goold}}, \bibinfo {author} {\bibfnamefont {T.}~\bibnamefont {Fogarty}},
  \bibinfo {author} {\bibfnamefont {N.}~\bibnamefont {Lo~Gullo}}, \bibinfo
  {author} {\bibfnamefont {M.}~\bibnamefont {Paternostro}}, \ and\ \bibinfo
  {author} {\bibfnamefont {T.}~\bibnamefont {Busch}},\ }\href {\doibase
  10.1103/PhysRevA.84.063632} {\bibfield  {journal} {\bibinfo  {journal} {Phys.
  Rev. A}\ }\textbf {\bibinfo {volume} {85}},\ \bibinfo {pages} {063632}
  (\bibinfo {year} {2011})}\BibitemShut {NoStop}%
\bibitem [{\citenamefont {Knap}\ \emph {et~al.}(2012)\citenamefont {Knap},
  \citenamefont {Shashi}, \citenamefont {Nishida}, \citenamefont {Imambekov},
  \citenamefont {Abanin},\ and\ \citenamefont {Demler}}]{knap2012tdi}%
  \BibitemOpen
  \bibfield  {author} {\bibinfo {author} {\bibfnamefont {M.}~\bibnamefont
  {Knap}}, \bibinfo {author} {\bibfnamefont {A.}~\bibnamefont {Shashi}},
  \bibinfo {author} {\bibfnamefont {Y.}~\bibnamefont {Nishida}}, \bibinfo
  {author} {\bibfnamefont {A.}~\bibnamefont {Imambekov}}, \bibinfo {author}
  {\bibfnamefont {D.~A.}\ \bibnamefont {Abanin}}, \ and\ \bibinfo {author}
  {\bibfnamefont {E.}~\bibnamefont {Demler}},\ }\href {\doibase
  10.1103/PhysRevX.2.041020} {\bibfield  {journal} {\bibinfo  {journal} {Phys.
  Rev. X}\ }\textbf {\bibinfo {volume} {2}},\ \bibinfo {pages} {041020}
  (\bibinfo {year} {2012})}\BibitemShut {NoStop}%
\bibitem [{\citenamefont {Chevy}(2006)}]{Chevy2006upd}%
  \BibitemOpen
  \bibfield  {author} {\bibinfo {author} {\bibfnamefont {F.}~\bibnamefont
  {Chevy}},\ }\href {\doibase 10.1103/PhysRevA.74.063628} {\bibfield  {journal}
  {\bibinfo  {journal} {Phys. Rev. A}\ }\textbf {\bibinfo {volume} {74}},\
  \bibinfo {pages} {063628} (\bibinfo {year} {2006})}\BibitemShut {NoStop}%
\bibitem [{\citenamefont {Prokof'ev}\ and\ \citenamefont
  {Svistunov}(2008)}]{Prokofev2008}%
  \BibitemOpen
  \bibfield  {author} {\bibinfo {author} {\bibfnamefont {N.}~\bibnamefont
  {Prokof'ev}}\ and\ \bibinfo {author} {\bibfnamefont {B.}~\bibnamefont
  {Svistunov}},\ }\href {\doibase 10.1103/PhysRevB.77.020408} {\bibfield
  {journal} {\bibinfo  {journal} {Phys. Rev. B}\ }\textbf {\bibinfo {volume}
  {77}},\ \bibinfo {pages} {020408} (\bibinfo {year} {2008})}\BibitemShut
  {NoStop}%
\bibitem [{\citenamefont {Loschmidt}(1876)}]{Loschmidt1876udz}%
  \BibitemOpen
  \bibfield  {author} {\bibinfo {author} {\bibfnamefont {J.}~\bibnamefont
  {Loschmidt}},\ }\href@noop {} {\bibfield  {journal} {\bibinfo  {journal}
  {Sitzungsberichte der Akademie der Wissenschaften, Wien}\ }\textbf {\bibinfo
  {volume} {73}},\ \bibinfo {pages} {128} (\bibinfo {year} {1876})}\BibitemShut
  {NoStop}%
\bibitem [{\citenamefont {Gurarie}\ and\ \citenamefont
  {Radzihovsky}(2007)}]{Gurarie2007}%
  \BibitemOpen
  \bibfield  {author} {\bibinfo {author} {\bibfnamefont {V.}~\bibnamefont
  {Gurarie}}\ and\ \bibinfo {author} {\bibfnamefont {L.}~\bibnamefont
  {Radzihovsky}},\ }\href {\doibase
  http://dx.doi.org/10.1016/j.aop.2006.10.009} {\bibfield  {journal} {\bibinfo
  {journal} {Annals of Physics}\ }\textbf {\bibinfo {volume} {322}},\ \bibinfo
  {pages} {2 } (\bibinfo {year} {2007})}\BibitemShut {NoStop}%
\bibitem [{\citenamefont {Petrov}(2004)}]{Petrov2004tbp}%
  \BibitemOpen
  \bibfield  {author} {\bibinfo {author} {\bibfnamefont {D.~S.}\ \bibnamefont
  {Petrov}},\ }\href {\doibase 10.1103/PhysRevLett.93.143201} {\bibfield
  {journal} {\bibinfo  {journal} {Phys. Rev. Lett.}\ }\textbf {\bibinfo
  {volume} {93}},\ \bibinfo {pages} {143201} (\bibinfo {year}
  {2004})}\BibitemShut {NoStop}%
\bibitem [{\citenamefont {McLachlan}(1964)}]{McLachlan64}%
  \BibitemOpen
  \bibfield  {author} {\bibinfo {author} {\bibfnamefont {A.~D.}\ \bibnamefont
  {McLachlan}},\ }\href {\doibase 10.1080/00268976400100041} {\bibfield
  {journal} {\bibinfo  {journal} {Molecular Physics}\ }\textbf {\bibinfo
  {volume} {8}},\ \bibinfo {pages} {39} (\bibinfo {year} {1964})}\BibitemShut
  {NoStop}%
\bibitem [{\citenamefont {Basile}\ and\ \citenamefont
  {Elser}(1995)}]{PhysRevE.51.5688}%
  \BibitemOpen
  \bibfield  {author} {\bibinfo {author} {\bibfnamefont {A.~G.}\ \bibnamefont
  {Basile}}\ and\ \bibinfo {author} {\bibfnamefont {V.}~\bibnamefont {Elser}},\
  }\href {\doibase 10.1103/PhysRevE.51.5688} {\bibfield  {journal} {\bibinfo
  {journal} {Phys. Rev. E}\ }\textbf {\bibinfo {volume} {51}},\ \bibinfo
  {pages} {5688} (\bibinfo {year} {1995})}\BibitemShut {NoStop}%
\bibitem [{\citenamefont {Combescot}\ \emph {et~al.}(2007)\citenamefont
  {Combescot}, \citenamefont {Recati}, \citenamefont {Lobo},\ and\
  \citenamefont {Chevy}}]{CombescotLoboChevy2007}%
  \BibitemOpen
  \bibfield  {author} {\bibinfo {author} {\bibfnamefont {R.}~\bibnamefont
  {Combescot}}, \bibinfo {author} {\bibfnamefont {A.}~\bibnamefont {Recati}},
  \bibinfo {author} {\bibfnamefont {C.}~\bibnamefont {Lobo}}, \ and\ \bibinfo
  {author} {\bibfnamefont {F.}~\bibnamefont {Chevy}},\ }\href {\doibase
  10.1103/PhysRevLett.98.180402} {\bibfield  {journal} {\bibinfo  {journal}
  {Phys. Rev. Lett.}\ }\textbf {\bibinfo {volume} {98}},\ \bibinfo {pages}
  {180402} (\bibinfo {year} {2007})}\BibitemShut {NoStop}%
\bibitem [{\citenamefont {Combescot}\ and\ \citenamefont
  {Giraud}(2008)}]{combescot2008nso}%
  \BibitemOpen
  \bibfield  {author} {\bibinfo {author} {\bibfnamefont {R.}~\bibnamefont
  {Combescot}}\ and\ \bibinfo {author} {\bibfnamefont {S.}~\bibnamefont
  {Giraud}},\ }\href@noop {} {\bibfield  {journal} {\bibinfo  {journal} {Phys.
  Rev. Lett.}\ }\textbf {\bibinfo {volume} {101}},\ \bibinfo {pages} {050404}
  (\bibinfo {year} {2008})}\BibitemShut {NoStop}%
\bibitem [{\citenamefont {Combescot}\ \emph {et~al.}(2009)\citenamefont
  {Combescot}, \citenamefont {Giraud},\ and\ \citenamefont
  {Leyronas}}]{combescot2009}%
  \BibitemOpen
  \bibfield  {author} {\bibinfo {author} {\bibfnamefont {R.}~\bibnamefont
  {Combescot}}, \bibinfo {author} {\bibfnamefont {S.}~\bibnamefont {Giraud}}, \
  and\ \bibinfo {author} {\bibfnamefont {X.}~\bibnamefont {Leyronas}},\ }\href
  {\doibase 0295-5075/88/60007} {\bibfield  {journal} {\bibinfo  {journal}
  {Europhys. Lett.}\ }\textbf {\bibinfo {volume} {88}},\ \bibinfo {pages}
  {60007} (\bibinfo {year} {2009})}\BibitemShut {NoStop}%
\bibitem [{\citenamefont {Punk}\ \emph {et~al.}(2009)\citenamefont {Punk},
  \citenamefont {Dumitrescu},\ and\ \citenamefont {Zwerger}}]{punk2009}%
  \BibitemOpen
  \bibfield  {author} {\bibinfo {author} {\bibfnamefont {M.}~\bibnamefont
  {Punk}}, \bibinfo {author} {\bibfnamefont {P.~T.}\ \bibnamefont
  {Dumitrescu}}, \ and\ \bibinfo {author} {\bibfnamefont {W.}~\bibnamefont
  {Zwerger}},\ }\href {\doibase 10.1103/PhysRevA.80.053605} {\bibfield
  {journal} {\bibinfo  {journal} {Phys. Rev. A}\ }\textbf {\bibinfo {volume}
  {80}},\ \bibinfo {eid} {053605} (\bibinfo {year} {2009})}\BibitemShut
  {NoStop}%
\bibitem [{\citenamefont {Mora}\ and\ \citenamefont
  {Chevy}(2009)}]{Mora2009gso}%
  \BibitemOpen
  \bibfield  {author} {\bibinfo {author} {\bibfnamefont {C.}~\bibnamefont
  {Mora}}\ and\ \bibinfo {author} {\bibfnamefont {F.}~\bibnamefont {Chevy}},\
  }\href@noop {} {\bibfield  {journal} {\bibinfo  {journal} {Phys. Rev. A}\
  }\textbf {\bibinfo {volume} {80}},\ \bibinfo {pages} {033607} (\bibinfo
  {year} {2009})}\BibitemShut {NoStop}%
\bibitem [{\citenamefont {Mathy}\ \emph {et~al.}(2011)\citenamefont {Mathy},
  \citenamefont {Parish},\ and\ \citenamefont {Huse}}]{Mathy2011tma}%
  \BibitemOpen
  \bibfield  {author} {\bibinfo {author} {\bibfnamefont {C.~J.~M.}\
  \bibnamefont {Mathy}}, \bibinfo {author} {\bibfnamefont {M.~M.}\ \bibnamefont
  {Parish}}, \ and\ \bibinfo {author} {\bibfnamefont {D.~A.}\ \bibnamefont
  {Huse}},\ }\href {\doibase 10.1103/PhysRevLett.106.166404} {\bibfield
  {journal} {\bibinfo  {journal} {Phys. Rev. Lett.}\ }\textbf {\bibinfo
  {volume} {106}},\ \bibinfo {pages} {166404} (\bibinfo {year}
  {2011})}\BibitemShut {NoStop}%
\bibitem [{\citenamefont {Cui}\ and\ \citenamefont {Zhai}(2010)}]{Cui2010}%
  \BibitemOpen
  \bibfield  {author} {\bibinfo {author} {\bibfnamefont {X.}~\bibnamefont
  {Cui}}\ and\ \bibinfo {author} {\bibfnamefont {H.}~\bibnamefont {Zhai}},\
  }\href {\doibase 10.1103/PhysRevA.81.041602} {\bibfield  {journal} {\bibinfo
  {journal} {Phys. Rev. A}\ }\textbf {\bibinfo {volume} {81}},\ \bibinfo
  {pages} {041602} (\bibinfo {year} {2010})}\BibitemShut {NoStop}%
\bibitem [{Note1()}]{Note1}%
  \BibitemOpen
  \bibinfo {note} {Here we have assumed that all the closed channel molecules
  have been converted into spin-up atoms at the time of the measurement, which
  is reasonable if there is a sufficient delay after the second rf
  pulse.}\BibitemShut {Stop}%
\bibitem [{\citenamefont {Trefzger}\ and\ \citenamefont
  {Castin}(2012)}]{Trefzger2012}%
  \BibitemOpen
  \bibfield  {author} {\bibinfo {author} {\bibfnamefont {C.}~\bibnamefont
  {Trefzger}}\ and\ \bibinfo {author} {\bibfnamefont {Y.}~\bibnamefont
  {Castin}},\ }\href {\doibase 10.1103/PhysRevA.85.053612} {\bibfield
  {journal} {\bibinfo  {journal} {Phys. Rev. A}\ }\textbf {\bibinfo {volume}
  {85}},\ \bibinfo {pages} {053612} (\bibinfo {year} {2012})}\BibitemShut
  {NoStop}%
\bibitem [{\citenamefont {{Goulko}}\ \emph {et~al.}()\citenamefont {{Goulko}},
  \citenamefont {{Mishchenko}}, \citenamefont {{Prokof'ev}},\ and\
  \citenamefont {{Svistunov}}}]{Goulko2016}%
  \BibitemOpen
  \bibfield  {author} {\bibinfo {author} {\bibfnamefont {O.}~\bibnamefont
  {{Goulko}}}, \bibinfo {author} {\bibfnamefont {A.~S.}\ \bibnamefont
  {{Mishchenko}}}, \bibinfo {author} {\bibfnamefont {N.}~\bibnamefont
  {{Prokof'ev}}}, \ and\ \bibinfo {author} {\bibfnamefont {B.}~\bibnamefont
  {{Svistunov}}},\ }\href@noop {} {\enquote {\bibinfo {title} {{Dark Continuum
  in the Spectral Function of the Resonant Fermi Polaron}},}\ }\bibinfo {note}
  {ArXiv:1603.06963}\BibitemShut {NoStop}%
\bibitem [{\citenamefont {Massignan}(2012)}]{Massignan2012}%
  \BibitemOpen
  \bibfield  {author} {\bibinfo {author} {\bibfnamefont {P.}~\bibnamefont
  {Massignan}},\ }\href {\doibase 10.1209/0295-5075/98/10012} {\bibfield
  {journal} {\bibinfo  {journal} {EPL (Europhysics Letters)}\ }\textbf
  {\bibinfo {volume} {98}},\ \bibinfo {pages} {10012} (\bibinfo {year}
  {2012})}\BibitemShut {NoStop}%
\bibitem [{\citenamefont {Schmidt}\ and\ \citenamefont
  {Enss}(2011)}]{Schmidt2011}%
  \BibitemOpen
  \bibfield  {author} {\bibinfo {author} {\bibfnamefont {R.}~\bibnamefont
  {Schmidt}}\ and\ \bibinfo {author} {\bibfnamefont {T.}~\bibnamefont {Enss}},\
  }\href {\doibase 10.1103/PhysRevA.83.063620} {\bibfield  {journal} {\bibinfo
  {journal} {Phys. Rev. A}\ }\textbf {\bibinfo {volume} {83}},\ \bibinfo
  {pages} {063620} (\bibinfo {year} {2011})}\BibitemShut {NoStop}%
\bibitem [{\citenamefont {Massignan}\ and\ \citenamefont
  {Bruun}(2011)}]{Massignan2011}%
  \BibitemOpen
  \bibfield  {author} {\bibinfo {author} {\bibfnamefont {P.}~\bibnamefont
  {Massignan}}\ and\ \bibinfo {author} {\bibfnamefont {G.~M.}\ \bibnamefont
  {Bruun}},\ }\href {\doibase 10.1140/epjd/e2011-20084-5} {\bibfield  {journal}
  {\bibinfo  {journal} {The European Physical Journal D}\ }\textbf {\bibinfo
  {volume} {65}},\ \bibinfo {pages} {83} (\bibinfo {year} {2011})}\BibitemShut
  {NoStop}%
\bibitem [{\citenamefont {Mahan}(1990)}]{mahan1990mpp}%
  \BibitemOpen
  \bibfield  {author} {\bibinfo {author} {\bibfnamefont {G.~D.}\ \bibnamefont
  {Mahan}},\ }\href@noop {} {\emph {\bibinfo {title} {Many-particle
  physics}}},\ Physics of solids and liquids\ (\bibinfo  {publisher} {Plenum},\
  \bibinfo {address} {New York, NY},\ \bibinfo {year} {1990})\BibitemShut
  {NoStop}%
\bibitem [{\citenamefont {Massignan}\ \emph {et~al.}(2013)\citenamefont
  {Massignan}, \citenamefont {Yu},\ and\ \citenamefont
  {Bruun}}]{Massignan2013}%
  \BibitemOpen
  \bibfield  {author} {\bibinfo {author} {\bibfnamefont {P.}~\bibnamefont
  {Massignan}}, \bibinfo {author} {\bibfnamefont {Z.}~\bibnamefont {Yu}}, \
  and\ \bibinfo {author} {\bibfnamefont {G.~M.}\ \bibnamefont {Bruun}},\ }\href
  {\doibase 10.1103/PhysRevLett.110.230401} {\bibfield  {journal} {\bibinfo
  {journal} {Phys. Rev. Lett.}\ }\textbf {\bibinfo {volume} {110}},\ \bibinfo
  {pages} {230401} (\bibinfo {year} {2013})}\BibitemShut {NoStop}%
\bibitem [{\citenamefont {Xu}\ \emph {et~al.}(2013)\citenamefont {Xu},
  \citenamefont {Gu},\ and\ \citenamefont {Mueller}}]{mueller2013}%
  \BibitemOpen
  \bibfield  {author} {\bibinfo {author} {\bibfnamefont {J.}~\bibnamefont
  {Xu}}, \bibinfo {author} {\bibfnamefont {Q.}~\bibnamefont {Gu}}, \ and\
  \bibinfo {author} {\bibfnamefont {E.~J.}\ \bibnamefont {Mueller}},\ }\href
  {\doibase 10.1103/PhysRevA.88.023604} {\bibfield  {journal} {\bibinfo
  {journal} {Phys. Rev. A}\ }\textbf {\bibinfo {volume} {88}},\ \bibinfo
  {pages} {023604} (\bibinfo {year} {2013})}\BibitemShut {NoStop}%
\bibitem [{\citenamefont {Tan}(2008)}]{Tan2008eoa}%
  \BibitemOpen
  \bibfield  {author} {\bibinfo {author} {\bibfnamefont {S.}~\bibnamefont
  {Tan}},\ }\href {\doibase 10.1016/j.aop.2008.03.004} {\bibfield  {journal}
  {\bibinfo  {journal} {Ann. Phys.}\ }\textbf {\bibinfo {volume} {323}},\
  \bibinfo {pages} {2952} (\bibinfo {year} {2008})}\BibitemShut {NoStop}%
\bibitem [{\citenamefont {Kartavtsev}\ and\ \citenamefont
  {Malykh}(2007)}]{Kartavtsev2007let}%
  \BibitemOpen
  \bibfield  {author} {\bibinfo {author} {\bibfnamefont {O.~I.}\ \bibnamefont
  {Kartavtsev}}\ and\ \bibinfo {author} {\bibfnamefont {A.~V.}\ \bibnamefont
  {Malykh}},\ }\href@noop {} {\bibfield  {journal} {\bibinfo  {journal} {J.
  Phys. B}\ }\textbf {\bibinfo {volume} {40}},\ \bibinfo {pages} {1429}
  (\bibinfo {year} {2007})}\BibitemShut {NoStop}%
\bibitem [{\citenamefont {Levinsen}\ \emph {et~al.}(2009)\citenamefont
  {Levinsen}, \citenamefont {Tiecke}, \citenamefont {Walraven},\ and\
  \citenamefont {Petrov}}]{Levinsen2009ads}%
  \BibitemOpen
  \bibfield  {author} {\bibinfo {author} {\bibfnamefont {J.}~\bibnamefont
  {Levinsen}}, \bibinfo {author} {\bibfnamefont {T.~G.}\ \bibnamefont
  {Tiecke}}, \bibinfo {author} {\bibfnamefont {J.~T.~M.}\ \bibnamefont
  {Walraven}}, \ and\ \bibinfo {author} {\bibfnamefont {D.~S.}\ \bibnamefont
  {Petrov}},\ }\href@noop {} {\bibfield  {journal} {\bibinfo  {journal} {Phys.
  Rev. Lett.}\ }\textbf {\bibinfo {volume} {103}},\ \bibinfo {pages} {153202}
  (\bibinfo {year} {2009})}\BibitemShut {NoStop}%
\bibitem [{\citenamefont {Efimov}(1973)}]{Efimov1973elo}%
  \BibitemOpen
  \bibfield  {author} {\bibinfo {author} {\bibfnamefont {V.}~\bibnamefont
  {Efimov}},\ }\href@noop {} {\bibfield  {journal} {\bibinfo  {journal} {Nucl.
  Phys. A}\ }\textbf {\bibinfo {volume} {210}},\ \bibinfo {pages} {157}
  (\bibinfo {year} {1973})}\BibitemShut {NoStop}%
\bibitem [{\citenamefont {Castin}\ \emph {et~al.}(2010)\citenamefont {Castin},
  \citenamefont {Mora},\ and\ \citenamefont {Pricoupenko}}]{Castin2010fbe}%
  \BibitemOpen
  \bibfield  {author} {\bibinfo {author} {\bibfnamefont {Y.}~\bibnamefont
  {Castin}}, \bibinfo {author} {\bibfnamefont {C.}~\bibnamefont {Mora}}, \ and\
  \bibinfo {author} {\bibfnamefont {L.}~\bibnamefont {Pricoupenko}},\
  }\href@noop {} {\bibfield  {journal} {\bibinfo  {journal} {Phys. Rev. Lett.}\
  }\textbf {\bibinfo {volume} {105}},\ \bibinfo {pages} {223201} (\bibinfo
  {year} {2010})}\BibitemShut {NoStop}%
\bibitem [{\citenamefont {Cetina}\ \emph {et~al.}(2015)\citenamefont {Cetina},
  \citenamefont {Jag}, \citenamefont {Lous}, \citenamefont {Walraven},
  \citenamefont {Grimm}, \citenamefont {Christensen},\ and\ \citenamefont
  {Bruun}}]{Cetina2015doi}%
  \BibitemOpen
  \bibfield  {author} {\bibinfo {author} {\bibfnamefont {M.}~\bibnamefont
  {Cetina}}, \bibinfo {author} {\bibfnamefont {M.}~\bibnamefont {Jag}},
  \bibinfo {author} {\bibfnamefont {R.~S.}\ \bibnamefont {Lous}}, \bibinfo
  {author} {\bibfnamefont {J.~T.~M.}\ \bibnamefont {Walraven}}, \bibinfo
  {author} {\bibfnamefont {R.}~\bibnamefont {Grimm}}, \bibinfo {author}
  {\bibfnamefont {R.~S.}\ \bibnamefont {Christensen}}, \ and\ \bibinfo {author}
  {\bibfnamefont {G.~M.}\ \bibnamefont {Bruun}},\ }\href {\doibase
  10.1103/PhysRevLett.115.135302} {\bibfield  {journal} {\bibinfo  {journal}
  {Phys. Rev. Lett.}\ }\textbf {\bibinfo {volume} {115}},\ \bibinfo {pages}
  {135302} (\bibinfo {year} {2015})}\BibitemShut {NoStop}%
\bibitem [{\citenamefont {Rosch}\ and\ \citenamefont
  {Kopp}(1995)}]{rosch1995hpi}%
  \BibitemOpen
  \bibfield  {author} {\bibinfo {author} {\bibfnamefont {A.}~\bibnamefont
  {Rosch}}\ and\ \bibinfo {author} {\bibfnamefont {T.}~\bibnamefont {Kopp}},\
  }\href@noop {} {\bibfield  {journal} {\bibinfo  {journal} {Phys. Rev. Lett.}\
  }\textbf {\bibinfo {volume} {75}},\ \bibinfo {pages} {1988} (\bibinfo {year}
  {1995})}\BibitemShut {NoStop}%
\bibitem [{\citenamefont {J\o{}rgensen}\ \emph {et~al.}(2016)\citenamefont
  {J\o{}rgensen}, \citenamefont {Wacker}, \citenamefont {Skalmstang},
  \citenamefont {Parish}, \citenamefont {Levinsen}, \citenamefont
  {Christensen}, \citenamefont {Bruun},\ and\ \citenamefont
  {Arlt}}]{Jorgensen2016}%
  \BibitemOpen
  \bibfield  {author} {\bibinfo {author} {\bibfnamefont {N.~B.}\ \bibnamefont
  {J\o{}rgensen}}, \bibinfo {author} {\bibfnamefont {L.}~\bibnamefont
  {Wacker}}, \bibinfo {author} {\bibfnamefont {K.~T.}\ \bibnamefont
  {Skalmstang}}, \bibinfo {author} {\bibfnamefont {M.~M.}\ \bibnamefont
  {Parish}}, \bibinfo {author} {\bibfnamefont {J.}~\bibnamefont {Levinsen}},
  \bibinfo {author} {\bibfnamefont {R.~S.}\ \bibnamefont {Christensen}},
  \bibinfo {author} {\bibfnamefont {G.~M.}\ \bibnamefont {Bruun}}, \ and\
  \bibinfo {author} {\bibfnamefont {J.~J.}\ \bibnamefont {Arlt}},\ }\href
  {\doibase 10.1103/PhysRevLett.117.055302} {\bibfield  {journal} {\bibinfo
  {journal} {Phys. Rev. Lett.}\ }\textbf {\bibinfo {volume} {117}},\ \bibinfo
  {pages} {055302} (\bibinfo {year} {2016})}\BibitemShut {NoStop}%
\bibitem [{\citenamefont {Nishida}(2015)}]{Nishida2015}%
  \BibitemOpen
  \bibfield  {author} {\bibinfo {author} {\bibfnamefont {Y.}~\bibnamefont
  {Nishida}},\ }\href {\doibase 10.1103/PhysRevLett.114.115302} {\bibfield
  {journal} {\bibinfo  {journal} {Phys. Rev. Lett.}\ }\textbf {\bibinfo
  {volume} {114}},\ \bibinfo {pages} {115302} (\bibinfo {year}
  {2015})}\BibitemShut {NoStop}%
\bibitem [{\citenamefont {Yi}\ and\ \citenamefont {Cui}(2015)}]{Wei2015}%
  \BibitemOpen
  \bibfield  {author} {\bibinfo {author} {\bibfnamefont {W.}~\bibnamefont
  {Yi}}\ and\ \bibinfo {author} {\bibfnamefont {X.}~\bibnamefont {Cui}},\
  }\href {\doibase 10.1103/PhysRevA.92.013620} {\bibfield  {journal} {\bibinfo
  {journal} {Phys. Rev. A}\ }\textbf {\bibinfo {volume} {92}},\ \bibinfo
  {pages} {013620} (\bibinfo {year} {2015})}\BibitemShut {NoStop}%
\bibitem [{\citenamefont {Makotyn}\ \emph {et~al.}(2014)\citenamefont
  {Makotyn}, \citenamefont {Klauss}, \citenamefont {Goldberger}, \citenamefont
  {Cornell},\ and\ \citenamefont {Jin}}]{Makotyn:2014fk}%
  \BibitemOpen
  \bibfield  {author} {\bibinfo {author} {\bibfnamefont {P.}~\bibnamefont
  {Makotyn}}, \bibinfo {author} {\bibfnamefont {C.~E.}\ \bibnamefont {Klauss}},
  \bibinfo {author} {\bibfnamefont {D.~L.}\ \bibnamefont {Goldberger}},
  \bibinfo {author} {\bibfnamefont {E.~A.}\ \bibnamefont {Cornell}}, \ and\
  \bibinfo {author} {\bibfnamefont {D.~S.}\ \bibnamefont {Jin}},\ }\href
  {http://dx.doi.org/10.1038/nphys2850} {\bibfield  {journal} {\bibinfo
  {journal} {Nat Phys}\ }\textbf {\bibinfo {volume} {10}},\ \bibinfo {pages}
  {116} (\bibinfo {year} {2014})}\BibitemShut {NoStop}%
\end{thebibliography}%



\end{document}